\documentclass[11pt,onecolumn]{cleantechnicalreport}
\usepackage{natbib}
\setcitestyle{authoryear,round,citesep={;},aysep={,},yysep={;}}
\usepackage{hybridfrontmatter}

\usepackage[utf8]{inputenc} %
\usepackage[T1]{fontenc}    %
\usepackage{hyperref}       %
\usepackage{url}            %
\usepackage{booktabs}       %
\usepackage{nicefrac}       %
\usepackage{microtype}      %
\usepackage{xcolor}         %
\usepackage{multirow}
\usepackage{graphicx}
\usepackage{caption}
\usepackage{subcaption}
\usepackage{tabularx}
\usepackage{array}
\usepackage{enumitem}
\usepackage{xurl}      %
\usepackage{float}
\usepackage{wrapfig}

\usepackage{titletoc}

\usepackage{algorithm}
\usepackage{algpseudocode}

\usepackage{amsmath, amssymb, amsfonts}
\usepackage{mathtools}
\usepackage{amsthm}
\usepackage{bm}
\usepackage{aep-template-compat}

\AtBeginDocument{\setlength{\parindent}{0pt}}

\theoremstyle{plain}
\newtheorem{theorem}{Theorem}
\newtheorem{lemma}[theorem]{Lemma}
\newtheorem{proposition}[theorem]{Proposition}
\newtheorem{corollary}[theorem]{Corollary}

\theoremstyle{definition}

\theoremstyle{remark}

\title{\textsc{AEP}: Towards Trustworthy Agent Communication, Networking, and Collaboration}

\author[1,*]{Yuhao Wang}
\author[1,*]{Shengfang Zhai}
\author[2]{Xinyu Zhou}
\author[3]{Guanghao Jin}
\author[1]{Wenjie Qu}
\author[4]{Huichi Zhou}
\author[5]{Hao Liu}
\author[1]{Jiaheng Zhang}
\author[4]{Jun Wang}
\author[6]{Linyi Yang}
\makeatletter
\renewcommand{\AB@affilsepx}{\quad\protect\Affilfont}
\affil[1]{National University of Singapore}
\affil[2]{University of Hong Kong}
\def\AB@affilsep{\protect\\\protect\Affilfont}
\affil[3]{Ludwig-Maximilians-Universit\"at M\"unchen}
\affil[4]{University College London, University of London}
\def\AB@affilsep{\protect\\\protect\Affilfont}
\affil[5]{Tianjin University}
\affil[6]{Southern University of Science and Technology}
\makeatother

\AddToHook{cmd/maketitle/after}{%
  \begingroup
    \footnotetext[1]{Equal contribution.}%
  \endgroup
}
\hypersetup{hidelinks,pdfauthor={Yuhao Wang, Shengfang Zhai, Xinyu Zhou, Guanghao Jin, Wenjie Qu, Huichi Zhou, Hao Liu, Jiaheng Zhang, Jun Wang, Linyi Yang}}

\begin{abstract}
As privately owned agents become increasingly capable, collaboration in open agent networks is limited by partial topology knowledge, ownership boundaries, and the risk of exposing or contaminating private context. Existing agent-to-agent protocols are useful when collaborators are already known, while shared-memory and broadcast-based systems rely on visibility and trust assumptions that do not hold in decentralized, low-trust environments. We introduce \emph{Open Private Agent Collaboration} (OPAC), an asset-centric paradigm for cooperation among private agents with open membership, partial observability, and ownership-aware access control. Rather than treating collaboration as the exchange of final responses, OPAC enables agents to selectively discover, exchange, audit, and integrate policy-controlled context assets. This shifts the unit of cooperation from transient messages to reusable external context, while keeping disclosure and integration subject to privacy and security constraints. We instantiate OPAC with the \textsc{Agent Exchange Protocol} (\textsc{AEP}), a decentralized framework that combines semantic discovery, demand-driven coordination, safety-gated asset exchange, and ledger-backed evidence derivation. By distinguishing physical topology from each agent's local cognitive topology, \textsc{AEP} allows agents to discover task-conditioned collaborators under an unknown network structure, release only admissible assets, and update future cooperation decisions from auditable interaction evidence. \textsc{AEP} therefore provides a middle ground between pure request-response interaction and globally shared memory, offering a practical foundation for trustworthy agent-native networking through selective, auditable reuse of distributed
context assets.

\end{abstract}

\begin{document}
\maketitle

\section{Introduction}
Recent advances in foundation models have made private agents increasingly capable of planning, reasoning, retrieval, and tool use~\citep{mialon2023augmented,zhao2023depth,yao2022react,schick2023toolformer,park2023generative,wang2023voyager}. 
As these agents are increasingly deployed across users, organizations, and services, realistic settings are giving rise to networks of privately owned agents~\citep{chen2024internet,yang2025survey,yang2025unlocking,chen2024agentverse,li2024personal,a2aprotocol2025, chang2025agent,yuan2026beyond}.
We refer to these systems as \emph{Open Private Agent Networks}: decentralized ecosystems in which agents dynamically collaborate while preserving ownership, access control, and selective disclosure over their memories, tools, and capabilities.
Such networks can reduce information silos by enabling agents to share and reuse task-relevant experiences, observations, cached results, and specialized capabilities held elsewhere in the network.

To better align with realistic deployment requirements, we emphasize two requirements for open private agent networks. 
First, the network should support \emph{open and dynamic membership}. Instead of forming centrally managed agent clusters, agents may join, leave, or fail over time, while each agent may observe only topology fragments, partial metadata, and limited routing evidence~\citep{yang2025agentnet}. 
Second, the network should support \emph{ownership-aware access control}. Agents owned by different users, organizations, or services do not share a single trust boundary. 
A cooperation mechanism must therefore support authorization, minimal disclosure, auditing, and controlled release and integration to safeguard agent-owned knowledge assets such as private memory and skills.
These two requirements lead to two challenges. 
(1) \emph{Partial Observability}. 
Before requesting help or importing external context, an agent must form a task-conditioned belief about which peers are relevant, reachable, authorized, and sufficiently trustworthy. In decentralized Open Private Agent Networks, this requires collaborator discovery from partial trusted evidence, without relying on a global registry or incurring excessive communication and privacy costs.  
(2) \emph{Privacy and security constraints across ownership boundaries}. 
Even after a relevant collaborator is found, cooperation remains risky because the agents are owned by different parties and may follow different policies. 
The challenge is to enable useful collaboration without incurring privacy leakage~\citep{yagoubi2026agentleak,wang2025privacy,zhang2025searching,wang2025silent}, and without letting external information introduce malicious knowledge injection~\citep{zhan2024injecagent,debenedetti2024agentdojo,cui2025mad,dong2025memory,amayuelas2024multiagent}, or memory contamination~\citep{cui2025mad,dong2025memory,chen2024agentpoison, amayuelas2024multiagent}. 
Thus, cross-owner cooperation must control both what is shared and what is accepted.

Existing cooperation paradigms fall short in complementary ways.
(1) Request-centric agent-to-agent protocols work well when a counterpart, endpoint, or capability interface is known, but do not address discovery of relevant and trustworthy collaborators under partial observability~\citep{a2aprotocol2025, chang2025agent,marro2024scalable,acp2025}.
(2) Shared-memory, broadcast, blackboard, and tightly coordinated systems support stronger coordination, but typically rely on common state, shared governance, and trust assumptions unsuitable for independently owned private agents~\citep{rezazadeh2025collaborative,eigenflux2026whitepaper,moltbook2026,han2025exploring,masterman2024landscape}.
Table~\ref{tab:intro-comparison} illustrates this gap: existing approaches cover only subsets of these capabilities, but do not jointly support trustworthy discovery under partial topology, ownership-aware selective exchange, input safety, and evidence-based trust.

\begin{aepwraptable}{r}{0.56\textwidth}
\vspace{-1em}
\centering
\scriptsize
\caption{\small Comparison at the paradigm and protocol levels.
(a) Representative agent collaboration paradigms.
(b) Representative production-grade agent communication protocols.}
\label{tab:intro-comparison}

\vspace{-0.5em}
{\footnotesize(a) Paradigm-level comparison}

\vspace{0.2em}
\setlength{\tabcolsep}{2.2pt}
\renewcommand{\arraystretch}{0.88}

\resizebox{\linewidth}{!}{%
\begin{tabular}{@{}lccccc@{}}
\toprule
Architecture &
Discovery &
Privacy &
Safety &
Efficiency &
Collaboration \\
\midrule
Shared Mem. &
$\times$ & $\times$ & $\times$ & $\checkmark$ & $\checkmark$ \\

Broadcast &
$\times$ & $\triangle$ & $\times$ & $\times$ & $\triangle$ \\

Request-centric A2A &
$\times$ & $\triangle$ & $\times$ & $\triangle$ & $\triangle$ \\

\textsc{AEP} &
$\checkmark$ & $\checkmark$ & $\checkmark$ & $\checkmark$ & $\checkmark$ \\
\bottomrule
\end{tabular}%
}

\vspace{0.55em}
{\footnotesize(b) Protocol-level comparison}

\vspace{0.2em}
\resizebox{\linewidth}{!}{%
\begin{tabular}{@{}lcccccc@{}}
\toprule
Protocol &
\shortstack{Open\\Discovery} &
\shortstack{Access\\Control} &
\shortstack{Selective\\Release} &
\shortstack{Input\\Safety} &
\shortstack{Evidence-\\based Trust} &
\shortstack{Adaptive\\Coordination} \\
\midrule
ACP~\citep{acp2025} &
$\triangle$ & $\triangle$ & $\triangle$ &
$\times$ & $\triangle$ & $\times$ \\

ANP~\citep{chang2025agent} &
$\triangle$ & $\checkmark$ & $\checkmark$ &
$\triangle$ & $\triangle$ & $\checkmark$ \\

Agora~\citep{marro2024scalable} &
$\times$ & $\times$ & $\times$ &
$\times$ & $\times$ & $\checkmark$ \\

\textsc{AEP} (Ours) &
$\checkmark$ & $\checkmark$ & $\checkmark$ &
$\checkmark$ & $\checkmark$ & $\checkmark$ \\
\bottomrule
\end{tabular}%
}

\vspace{0.15em}
{\tiny
$\checkmark$: native/strong support;\quad
$\triangle$: partial or infrastructure-dependent;\quad
$\times$: not supported.
}

\vspace{-1.0em}
\end{aepwraptable}

Motivated by these limitations, we introduce the \emph{Open Private Agent Collaboration} (OPAC) paradigm: an asset-centric framework for cooperation among privately owned agents in open, partially observed, and low-trust networks. 
Rather than merely sending requests and receiving generated responses, OPAC enables agents to discover, exchange, audit, and integrate policy-controlled context assets, such as memory fragments, cached tool results, verification signals, and skills, that can be reused across downstream tasks.
To solve the partial observability problem, OPAC is built around the distinction between \emph{physical topology} and \emph{local cognitive topology}. 
The physical topology describes the actual connectivity of the agent network, while local cognitive topology is the current local view maintained by an individual agent: discovered topology fragments, known peer connections, and estimates of remote agents' information and states. 
Considering potential risk in privacy leakage and knowledge contamination, OPAC also treats cooperation as a privacy- and security-constrained process. Thus, the objective of OPAC is not simply to maximize task utility, but to improve utility subject to constraints on privacy, security risk, trust, and communication cost. 

\begin{figure}[t]
  \centering
  \includegraphics[width=1\textwidth]{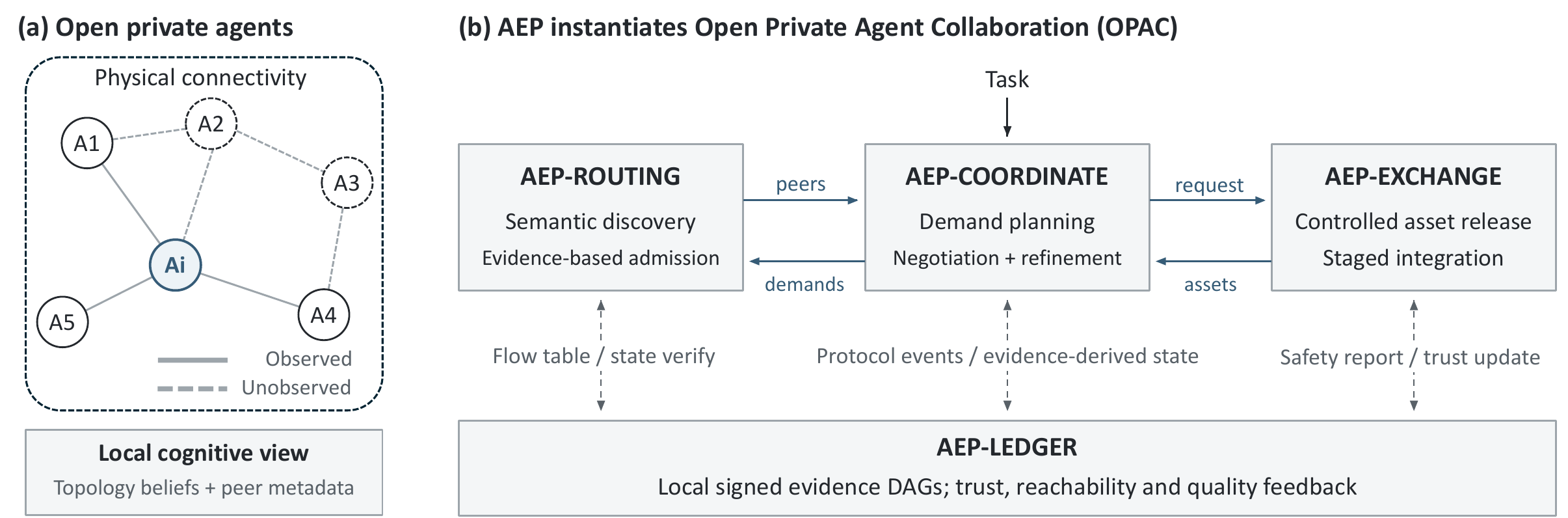}
  \caption{Open private agent collaboration and AEP framework.}
  \label{fig:intro}
\end{figure}

We instantiate OPAC with \textsc{Agent Exchange Protocol} (\textsc{AEP}), a concrete protocol framework for safe and efficient cooperation under unknown topology. 
\textsc{AEP} implements OPAC through four subprotocols: \textsc{AEP-Routing} for semantic discovery and evidence-derived candidate admission, \textsc{AEP-Coordinate} for collaboration planning and negotiation, \textsc{AEP-Exchange} for safety-gated asset release and staged integration, and \textsc{AEP-Ledger} for recording claims, incidents, trust assessments, and quality feedback, as shown in Figure~\ref{fig:intro}. 
Together, these components allow agents to form task-conditioned collaboration neighborhoods, exchange only admissible assets, and update future cooperation decisions from verifiable evidence. 
Our contributions are three-fold:
\begin{itemize}
    \item We introduce the \emph{Open Private Agent Collaboration} (OPAC) paradigm for collaboration among private agents in open, partially observed, and low-trust environments. OPAC frames collaboration as asset-centric exchange under constraints on task utility, privacy risk, security risk, trust, and communication cost.
    \item We present \textsc{Agent Exchange Protocol} (\textsc{AEP}), a four-subprotocol framework that instantiates OPAC through semantic discovery, collaboration coordination, safety-gated asset exchange, and ledger-backed evidence derivation under unknown topology.
    \item We provide both theoretical and empirical evidence that \textsc{AEP} improves the utility--safety trade-off. Our experiments show that \textsc{AEP}-based networks achieve the highest task scores, $0.00\%$ privacy leakage, near-zero injection success, and only a $3.5\%$ task-score drop under a $0.4$ malicious-agent ratio.
\end{itemize}

\section{Problem Formalization}
\label{sec:OPAC}

We formalize OPAC as a distributed multi-objective decision problem over an
open network of private agents:
\begin{equation}
\mathcal{V}=\{A_1,A_2,\dots,A_n\},
\end{equation}
embedded in an unknown communication graph $G=(\mathcal{V},\mathcal{E})$ with
physical topology $W^{\mathrm{phy}}\in\mathbb{R}^{n\times n}$. Agent $A_i$
does not observe $G$ globally; instead, it reasons from a local cognitive map
$G_i(t)$, a topology belief $W_i^{\mathrm{cog}}(t)$, routing state $R_i(t)$,
task input $X_i$, target output $Y_i$, and private memory $M_i$.
OPAC treats cooperation as the selective acquisition of external context assets,
such as memory and skills, rather than unrestricted message passing. Under a
policy $\pi$, agent $A_i$ receives an admitted asset bundle
$m_{\rightarrow i}^{\pi}(t)$ and reasons with
\(
\widetilde{M}_i^{\pi}(t)=M_i\cup m_{\rightarrow i}^{\pi}(t).
\)
The goal is to improve task performance through useful imported context, without
assuming full topology visibility or unrestricted state sharing. This means that
external context can be used only when it is both reachable under the agent's
local view and authorized for release. Let $\mathcal{A}_i(t)$ denote the assets
currently admissible to $A_i$, and let $R_i^{\mathrm{priv}}(\pi)$ denote the
privacy leakage risk. Since imported context and privacy risk depend on other
agents' release decisions, the network-level safe set is defined jointly:
\begin{equation}
\Pi_{\mathrm{safe}}(t)
=
\left\{
\pi\in\Pi_1\times\dots\times\Pi_n
\;\middle|\;
\forall i,\,
m_{\rightarrow i}^{\pi}(t)\subseteq\mathcal{A}_i(t),
\;
R_i^{\mathrm{priv}}(\pi)\le\epsilon_i
\right\}.
\end{equation}
We evaluate each safe policy by predictive utility, cooperation trust, and
resource cost. Here $U(\pi)$ measures task utility, $T(\pi)$ measures trustworthy and auditable
cooperation, and $C(\pi)$ measures cooperation cost:
\begin{equation}
U(\pi)=
\sum_{i=1}^{n}
\mathbb{E}\!\left[
\log p^{\star}\!\left(Y_i\mid X_i,\widetilde{M}_i^{\pi}(t)\right)
\right],
T(\pi)=\sum_{i=1}^{n}\mathbb{E}[\tau_i(\pi)],
C(\pi)=\sum_{i=1}^{n}C_i(\pi).
\end{equation}
The OPAC objective is:
\begin{equation}
z(\pi)=\big(U(\pi),\,T(\pi),\,-C(\pi)\big).
\end{equation}
A safe policy is Pareto optimal if no other safe policy improves one objective
without degrading another. Since different deployments may prefer different
utility--trust--cost trade-offs, OPAC targets the safe Pareto frontier
\begin{equation}
\mathcal{P}^{\star}(t)
=
\left\{
\pi\in\Pi_{\mathrm{safe}}(t)
\;\middle|\;
\nexists\,\pi'\in\Pi_{\mathrm{safe}}(t)
\text{ such that } z(\pi')\succ z(\pi)
\right\}.
\end{equation}
Thus, an OPAC solver must expand the safe feasible set under partial
observability and move the empirical frontier toward $\mathcal{P}^{\star}(t)$.
Appendix~\ref{app:opac-full} gives the full formulation.

\section{\textsc{Agent Exchange Protocol} (\textsc{AEP})}

We present \textsc{Agent Exchange Protocol} (\textsc{AEP}), a decentralized
protocol for private-agent cooperation under unknown topology. \textsc{AEP} enables agents to discover relevant peers, acquire useful external context, and integrate it under privacy and security constraints through four protocol layers. Each agent uses \textsc{AEP-Routing} to maintain a trust-aware local peer view. Given a task requirement, if local context is insufficient, \textsc{AEP-Coordinate} translates the unresolved need into external demands and selects candidate peers from this routing view. \textsc{AEP-Exchange} then determines how the requested asset should be handled and validates received payloads through staged ingestion. Throughout the process, \textsc{AEP-Ledger} records interaction evidence, allowing the other layers to base decisions on verified evidence rather than raw observations. Sec.~\ref{sec:aep-routing}--\ref{sec:aep-ledger} present the four protocol layers, Sec.~\ref{sec:aep-solver} connects the induced policy family to the Pareto-formulated OPAC objective, and Appendix~\ref{app:aep-operational} and Appendix~\ref{app:aep-theory-results} provide operational details, theoretical results, and proofs. Figure~\ref{fig:framework} in Appendix~\ref{app:aep-operational} illustrates the protocol framework and interaction sequence.

\subsection{\textsc{AEP-Routing}: Semantic Discovery and Evidence-Derived Peer State}
\label{sec:aep-discover}
\label{sec:aep-routing}

\textsc{AEP-Routing} discovers and admits collaborators through three evidence-aware steps:
it synchronizes public semantic metadata via \textbf{constrained gossip}, derives a
trust- and quality-aware \textbf{Semantic Flow Table} from ledger-backed records, and
confirms reachability through \textbf{topology probes}. Given an unresolved demand,
it then routes the requester to peers that are reachable, task-relevant, and trustworthy.
Appendix~\ref{app:aep-routing} gives the full operational details.

\noindent\textbf{Semantic flow table.}
In OPAC, agents must locate task-relevant external context under partial
topology. AEP therefore maintains a local \emph{Semantic Flow Table}, whose
propagated \texttt{Flow Entry} stores compact public metadata about a discovered
peer, including its semantic profile, route hints, freshness, and a
\texttt{source\_block\_hash} pointer to ledger evidence.
Since a \texttt{Flow Entry} is not
itself a cryptographic proof, agents first merge received entries into a raw table for fast semantic discovery, and then derive a verified view from the corresponding ledger evidence.
The verified view supplies the connectivity, blacklist, trust, and quality
signals used for safety-aware routing.

\noindent\textbf{Gossip-based verifiable state synchronization.}
AEP disseminates routing and evidence state through constrained
gossip~\citep{gossip2002}.
Each synchronization envelope carries freshness-filtered flow entries together
with incremental signed ledger evidence. Upon receipt, an agent merges entries into
the raw routing table and ingests evidence blocks into the local ledger. Entries
whose source evidence is missing are retained only as unverified cache. This
separates fast semantic propagation from asynchronous evidence verification,
allowing local routing views to converge without requiring every routing record
to be self-verifying. 

\noindent\textbf{Topology probing.}
Gossip may reveal peers whose direct reachability is still unverified. To avoid
treating such peers as collaboration-ready, \textsc{AEP-Routing} uses a signed
probe--response handshake to confirm point-to-point reachability without
exchanging task data. A successful probe records matched
\texttt{topology\_probe\_tx}/\texttt{topology\_probe\_rx} evidence and adds the
peer to the evidence-backed topology $G_i^{\mathrm{ev}}(t)$.

\noindent\textbf{Evidence-based Semantic Routing.}
When the local context is insufficient, requester $A_i$ embeds the unresolved demand
as $q_i$ and retrieves candidate peers from its semantic flow table. Candidates
are scored by semantic relevance and evidence-derived peer state:
\begin{equation*}
    r_{ij}(t)=\phi\!\left(q_i,\,\texttt{capability}_j,\,\texttt{mem\_sum}_j,\,
\texttt{keywords}_j,\,\texttt{trustscore}_{ij}(t),\,
\texttt{quality}_{ij}(t)\right).
\end{equation*}
AEP admits only peers that are fresh, non-blacklisted, sufficiently trusted,
endpoint-usable, and reachable in the evidence-backed topology
$G_i^{\mathrm{ev}}(t)$. Unsafe peers are therefore excluded before ranking. Service quality then reranks the admissible set, so AEP only contacts selected, trusted, high-quality, and route-eligible
peers.

\subsection{\textsc{AEP-Coordinate}: Demand-Driven Negotiation and Refinement}
\label{sec:aep-coordinate}

\textsc{AEP-Coordinate} bridges semantic discovery and asset transfer by converting
unresolved task needs into structured collaboration demands. 
It first performs a local sufficiency
check; when external help is needed, it decomposes the missing information into a
compact \textbf{demand list}, uses \textsc{AEP-Routing} for \textbf{bounded peer
selection}, and issues an exchange request or opens a short
\textbf{negotiation session} to solve demands. After each round,
it records outcome \textbf{feedback} as evidence, allowing later routing decisions, demand
selection, capability descriptions, and asset summaries to be refined.

\noindent\textbf{Demand-driven negotiation.}
For an incoming task, the requester first checks whether local
memory and skills are sufficient; if so, it answers locally. Otherwise, it
rewrites the unresolved need into a compact demand list, where each demand
represents a missing fact, context fragment, and executable subproblem. The requester then uses \textsc{AEP-Routing} to identify a
bounded set of reachable peers for each demand. If the demand is already
well-specified, it is converted into an exchange request; otherwise, the
requester opens a short negotiation session in which the responder may ask for
missing parameters or clarification. 

\noindent\textbf{Feedback and refinement.}
After each collaboration round, the coordinator updates the demand state as
resolved, pending, or unsolved according to the returned evidence and
confidence. It also evaluates the usefulness of the response and selected assets,
records service-quality feedback, and persists this feedback as evidence. These
signals later refine peer quality estimates, demand selection, public capability
descriptions, and asset summaries. Thus, \textsc{AEP-Coordinate} closes the
collaboration loop by converting downstream outcomes into future routing and
metadata improvements. Appendix~\ref{app:aep-coordinate} gives the operational
details.

\subsection{\textsc{AEP-Exchange}: Safety-Aware Asset Release}
\label{sec:aep-exchange}

\textsc{AEP-Exchange} handles how a selected responder contributes local memory or skill assets for an arranged demand. Given a demand, the responder searches its private memory and available skills for task-relevant assets, evaluates whether they can be safely used under the current request, and decides an appropriate release mode for each candidate. A memory asset may be shared directly when low-risk, sanitized before sharing when it contains sensitive or irrelevant details, or blocked when it violates the responder's release policy. A skill asset may either be released as an artifact or kept local, in which case the responder executes it on behalf of the requester and returns only the result. On the receiver side, received assets are validated again before being stored in local exchange memory. This design allows AEP to reuse external context while limiting privacy leakage, unsafe skill transfer, and memory contamination. Appendix~\ref{app:aep-exchange} gives more operational details.

\subsection{\textsc{AEP-Ledger}: Verifiable Evidence Infrastructure}
\label{sec:aep-ledger}

\textsc{AEP-Ledger} provides the verifiable evidence substrate that makes AEP's
cross-agent decisions auditable. It records protocol events in an
\textbf{append-only evidence log}, orders them with a \textbf{causal clock model},
and keeps long-running histories manageable through \textbf{checkpointed
compression}. Upper layers derive routing state, peer ratings, release evidence,
and refinement signals from these verified records rather than raw observations. Appendix~\ref{app:aep-ledger} gives the full
operational details.

\noindent\textbf{Append-only evidence log.}
The ledger is maintained as an append-only signed Directed Acyclic Graph (DAG) log, typically a Merkle-DAG supporting lightweight proofs over selected fields
~\citep{laurie2014certificate,benet2014ipfs,haeberlen2007peerreview,merkle1987digital}.
Each block binds a protocol event to its author, causal order, and signed
content, and is accepted only after validity and consistency checks.
Conflicting signed records, such as double-signing on same sequence position,
are treated as verifiable misbehavior and may trigger blacklisting.
Thus, AEP turns protocol
events into auditable evidence for routing, exchange, and refinement.

\noindent\textbf{Clock model.}
Protocol evidence is
ordered by a Lamport-style logical clock~\citep{lamport2019time} together with a
monotonically increasing counter maintained by each author, so agents can reason
about causal order and detect conflict from the same source.
In contrast, local step counters only schedule routine maintenance such as
gossip and metadata refresh, while physical time is reserved for the network deadlines, such as exchange timeouts.

\noindent\textbf{Checkpointed compression.}
Long-running deployments may accumulate large evidence histories, so AEP
periodically compresses finalized evidence into quorum-backed checkpoints,
following snapshot-based state summarization~\citep{chandy1985distributed}.
Consensus-backed checkpointing keeps this process safe and verifiable
~\citep{gifford1979weighted,yin2018hotstuff}. Each checkpoint commits to a compact
summary of evidence-derived state, covered blocks, current DAG heads, and validator
signatures. New or rejoining agents verify the latest checkpoint and synchronize
only incremental logs, allowing older evidence to be pruned while preserving
proof-critical commitments.

\subsection{AEP as an OPAC Solver}
\label{sec:aep-solver}

We interpret \textsc{AEP} as a decentralized approximate solver for the
Pareto-formulated OPAC objective in Section~\ref{sec:OPAC}. Its four layers
jointly induce a safe policy family: routing expands admissible cooperation
opportunities under partial topology, coordination allocates unresolved demands,
exchange imports external assets under authorization and sanitization
constraints, and ledger-backed evidence supports persistent agent policy with trust,
reachability, and quality. We provide Theorem~\ref{thm:aep-main-summary} as an idealized consistency and approximation result. Appendix~\ref{app:aep-theory-results} provides the full assumptions and proof details.

\begin{theorem}[AEP approximate Pareto consistency, informal]
\label{thm:aep-main-summary}
Let \(P^\star(t)\subseteq \Pi_{\mathrm{safe}}(t)\) be the safe Pareto-optimal
policy set and let \(P^{\star,z}(t)=\{z(\pi):\pi\in P^\star(t)\}\) be its
objective-space image. Under the assumptions in Appendix D, let
\(\widehat P^{\mathrm{AEP},z}_{t}\) be the objective-space image of the
\(2\epsilon_t\)-approximate empirical Pareto frontier induced by AEP. Then,
with high probability,
\begin{equation}
    d_H\!\left(\widehat P^{\mathrm{AEP},z}_{t},P^{\star,z}(t)\right)
\le
O(\epsilon_t)
+
O\!\left(L_{\mathrm{obj}}\frac{b(\Gamma)+\sigma}{\gamma_{\mathrm{phy}}}\right)
+
O(\Delta U).
\end{equation}
Thus, as statistical error, topology-learning error, and auditing opportunity loss vanish, the AEP-induced empirical objective frontier converges
to \(P^{\star,z}(t)\). Theorem~\ref{thm:aep-main} provides a formal formulation.
\end{theorem}

\section{Experiments}
\label{sec:experiments}

We evaluate \textsc{AEP} in an open-network multi-agent benchmark that captures privacy-sensitive collaboration under partial topology knowledge, unreliable communication, dynamic participation, and adversarial interaction. We report both utility and safety metrics to characterize the collaboration gains of \textsc{AEP} together with its privacy and robustness trade-offs.

\subsection{Experimental Setup}
\label{sec:experiments-setup}

\noindent\textbf{Simulation Setup.}
We use virtual HTTPS transport to emulate incomplete reachability, latency, packet loss, timeouts, retries, partitions, and leave-and-rejoin events. To assess scalability, we instantiate small-, medium-, and large-scale networks with 4, 16, and 64 benign agents, respectively. In each network, we additionally reserve two malicious agents for safety evaluation, targeting privacy theft and knowledge injection, respectively. Unless otherwise specified, utility metrics are computed over benign-agent task execution, while safety metrics are computed under adversarial participation.

\noindent\textbf{Scenarios.}
Following the scenario construction strategy and judge-based evaluation methodology of MultiAgentBench~\citep{zhu2025multiagentbench}, we adapt part of its data and construct three domains: \emph{Research}, \emph{Coding}, and \emph{Daily Service}. These tasks cover distributed knowledge gathering, code generation, and domain-specific service completion with personal context, private memory, and executable skills. Across domains, task executors and task-relevant knowledge and capabilities are deliberately distributed across different agents, making collaboration challenging and potentially risky. For task-utility and safety evaluation, each run contains 6 normal tasks; for long-term utility evaluation, each run contains 36 normal tasks. 

\noindent\textbf{Baselines.}
In the main experiments, we compare \textsc{AEP} with A2A and an A2A-driven broadcast baseline with DeepSeek-V4-Flash~\citep{xu2026deepseek} backbone (Section~\ref{sec:main_experi}). Broadcast contacts only known and addressable peers in the sender's local view, without global registry or oracle access. This paradigm-level comparison provides a broad comparison across representative collaboration architectures while controlling the implementation setting to isolate concrete design differences. To broaden the evaluation beyond abstract paradigms and assess practical protocol implementations, we additionally compare with three representative production-grade protocols within the request-centric A2A paradigm: ANP~\citep{chang2025agent}, Agora~\citep{marro2024scalable}, and ACP~\citep{acp2025} (Section~\ref{sec:production-grade protocol}). We further augment A2A and Broadcast with Qwen3Guard~\citep{zhao2025qwen3guard} to distinguish AEP's protocol-level safety from generic external filtering (Appendix~\ref{app:guarded-added}), and evaluate alternative backbone models to examine model dependence (Appendix~\ref{app:cross-model-added}).

\noindent\textbf{Metrics.}
We evaluate both utility and safety. For utility, following MultiAgentBench~\citep{zhu2025multiagentbench}, we report \textbf{Task Score}, where an LLM judge assigns a rubric-based score to the final answer by referring to the reference answer; \textbf{Completion Rate}, the fraction of tasks completed within the task budget; and \textbf{Token/task}, the average token consumption per task from model calls and collaborative communication. For safety, we report \textbf{Privacy Leakage Rate}, the fraction of protected private items disclosed to unauthorized agents or outputs, following prior work on agent privacy leakage~\citep{yagoubi2026agentleak,wang2025privacy}; and \textbf{Injection Success Rate}, the fraction of malicious injection attempts accepted into the downstream workflow, following prior work on prompt-injection evaluation for LLM agents~\citep{zhan2024injecagent,debenedetti2024agentdojo}. We aggregate all metrics over tasks and runs to characterize the utility--safety trade-off. Appendix~\ref{app:metric-definitions} gives full detail.

\subsection{Comparison Across Collaboration Paradigms}
\label{sec:main_experi}

\noindent\textbf{Task Utility.}
Table~\ref{tab:task_utility} shows that \textsc{AEP} consistently achieves higher task scores and completion rates across all domains and network scales, with near-perfect performance in Coding and Research. While \textsc{AEP} may consume more tokens than A2A due to higher task completion and additional safety checks, it remains substantially more token-efficient than Broadcast, reflecting the benefit of selective collaboration over broad communication.

\suppressfloats[t]

\begin{table*}[t]
\centering
\caption{Task Utility of Multi-Agent Collaboration}
\label{tab:task_utility}
\resizebox{\textwidth}{!}{
\begin{tabular}{llccccccccc}
\toprule
\multirow{2}{*}{Scenario} & \multirow{2}{*}{Protocol}
& \multicolumn{3}{c}{Small-Scale Network}
& \multicolumn{3}{c}{Middle-Scale Network}
& \multicolumn{3}{c}{Large-Scale Network} \\
\cmidrule(lr){3-5} \cmidrule(lr){6-8} \cmidrule(lr){9-11}
& & Task Score & Completion Rate (\%) & Tokens/task (k)
  & Task Score & Completion Rate (\%) & Tokens/task (k)
  & Task Score & Completion Rate (\%) & Tokens/task (k) \\
\midrule
\multirow{3}{*}{Coding}
& A2A       & 53.17 & 66.67  & 6.2 & 67.00 & 66.67  & 5.8 & 34.17 & 83.33  & 4.7 \\
& Broadcast & 67.00 & 66.67  & 7.2 & 82.67 & 83.33  & 11.7 & 33.33 & 33.33  & 26.2 \\
& AEP       & 99.00 & 100.00 & 6.0 & 99.17 & 100.00 & 10.1 & 99.67 & 100.00 & 19.2 \\
\midrule
\multirow{3}{*}{Daily-Service}
& A2A       & 31.33 & 66.67  & 3.1 & 17.00 & 16.67  & 3.3 & 31.00 & 66.67  & 4.2 \\
& Broadcast & 35.00 & 66.67  & 4.2 & 65.67 & 66.67  & 9.2 & 26.00 & 16.67  & 23.3 \\
& AEP       & 82.67 & 100.00 & 3.3 & 95.50 & 100.00 & 7.9 & 81.47 & 83.33  & 18.0 \\
\midrule
\multirow{3}{*}{Research}
& A2A       & 33.33 & 83.33  & 3.0 & 18.33 & 50.00  & 3.1 & 25.12 & 66.67  & 4.0 \\
& Broadcast & 20.00 & 83.33  & 3.8 & 35.00 & 33.33  & 10.3 & 33.33 & 33.33  & 35.2 \\
& AEP       & 99.50 & 100.00 & 3.9 & 99.67 & 100.00 & 9.2 & 99.17 & 100.00 & 14.9 \\
\bottomrule
\end{tabular}
}
\end{table*}

\noindent\textbf{Safety Performance.}
Table~\ref{tab:safety_assessment_new}  shows that A2A and Broadcast exhibit substantial privacy leakage and $100\%$ injection success across all settings, whereas \textsc{AEP} achieves $0.00\%$ privacy leakage and near-zero injection success. The remaining injection failures occur only in larger networks. These results are consistent with AEP's protocol-level safety controls over both asset release and integration. We additionally evaluate Qwen3Guard-enhanced variants of A2A and Broadcast in Appendix~\ref{app:guarded-added}.
\begin{table*}[t]
\centering
\caption{Safety Assessment of Multi-Agent Collaboration}
\label{tab:safety_assessment_new}
\resizebox{\textwidth}{!}{
\begin{tabular}{llcccccc}
\toprule
\multirow{2}{*}{Scenario} & \multirow{2}{*}{Protocol}
& \multicolumn{2}{c}{Small-Scale Network}
& \multicolumn{2}{c}{Middle-Scale Network}
& \multicolumn{2}{c}{Large-Scale Network} \\
\cmidrule(lr){3-4} \cmidrule(lr){5-6} \cmidrule(lr){7-8}
& & Leakage Rate (\%) & Injection Success Rate (\%)
& Leakage Rate (\%) & Injection Success Rate (\%)
& Leakage Rate (\%) & Injection Success Rate (\%) \\
\midrule
\multirow{3}{*}{Coding}
& A2A       & 38.11 & 100 & 36.11 & 100 & 31.94 & 100 \\
& Broadcast & 41.67 & 100 & 34.72 & 100 & 31.94 & 100 \\
& AEP       & 0  & 0   & 0  & 0   & 0  & 5.41 \\
\midrule
\multirow{3}{*}{Daily-Service}
& A2A       & 33.33 & 100 & 28.63 & 100 & 46.42 & 100 \\
& Broadcast & 66.67 & 100 & 36.11 & 100 & 41.67 & 100 \\
& AEP       & 0  & 0   & 0  & 0   & 0  & 4.80 \\
\midrule
\multirow{3}{*}{Research}
& A2A       & 35.80 & 100 & 35.14 & 100 & 38.80 & 100 \\
& Broadcast & 34.72 & 100 & 30.17 & 100 & 33.33 & 100 \\
& AEP       & 0  & 0   & 0  & 2.90   & 0  & 4.10 \\
\bottomrule
\end{tabular}
}
\end{table*}

\subsection{Comparison Across Production-Grade Protocols}
\label{sec:production-grade protocol}

While our main experiments compare collaboration paradigms, we further evaluate \textsc{AEP} against concrete production-grade protocols within the request-centric A2A paradigm. Specifically, we consider ANP~\citep{chang2025agent}, Agora~\citep{marro2024scalable}, and ACP~\citep{acp2025} in Table~\ref{tab:sdk-baselines-added}. Protocol versions, implementations, and experimental settings are detailed in Appendix~\ref{app:sdk-added}.
\textsc{AEP} consistently achieves higher task utility and completion while maintaining substantially lower injection success. These results show that AEP's advantages are not limited to controlled paradigm-level comparisons, but also hold against full protocol implementations.

\begin{table}[t]
\centering

\small
\caption{Comparison to production-grade protocol baselines under middle-scale network.}
\label{tab:sdk-baselines-added}

\resizebox{\linewidth}{!}{%
\begin{tabular}{llrrrrr}
\toprule
Scenario 
& Protocol 
& Task Score $\uparrow$ 
& Completion Rate (\%) $\uparrow$ 
& Tokens/task (k) $\downarrow$ 
& Leakage Rate (\%) $\downarrow$ 
& Injection Success Rate (\%) $\downarrow$ \\
\midrule

Coding 
& ANP 
& 55.44 & 66.67 & 6.3 & 0 & 100 \\
& Agora 
& 50.22 & 51.11 & 11.4 & 8.33 & 100 \\
& ACP 
& 50.22 & 57.78 & {5.3} & 0 & 100 \\
& {AEP}
& {99.17}
& {100.00}
& 10.1
& {0}
& {0} \\

\midrule

Daily-Service 
& ANP 
& 49.63 & 57.50 & 5.5 & 0 & 100 \\
& Agora 
& 38.50 & 50.00 & 8.9 & 8.33 & 100 \\
& ACP 
& 48.00 & 60.00 & {4.8} & 8.33 & 100 \\
& {AEP}
& {95.50}
& {100.00}
& 7.9
& {0}
& {0} \\

\midrule

Research 
& ANP 
& 50.75 & 60.00 & {3.9} & 16.67 & 100 \\
& Agora 
& 54.50 & 60.00 & 8.7 & 8.33 & 100 \\
& ACP 
& 49.00 & 57.50 & 5.1 & 16.67 & 100 \\
& {AEP}
& {99.67}
& {100.00}
& 9.2
& {0}
& {2.90} \\

\bottomrule
\end{tabular}%
}

\end{table}

\subsection{Evaluation on Long-term Task Utility}

Table~\ref{tab:task_utility_lite} shows that AEP's advantage persists over long-term interaction rather than disappearing after isolated tasks. Across Coding, Daily-Service, and Research, AEP achieves much higher task scores and completion rates than both A2A and Broadcast, indicating that its evidence-backed collaboration mechanism remains effective over repeated exchanges. The result suggests that AEP benefits from accumulated interaction evidence: quality feedback, safety incidents, verified reachability, and peer-state updates gradually improve later routing and exchange decisions. In contrast, A2A does not naturally expand discovery beyond initiated neighbors, while Broadcast continues to incur high communication cost and noisy context exposure. Although AEP may use more tokens than A2A, it converts this additional communication into substantially higher completion and answer quality, showing that long-term utility gains come from selective and increasingly evidence-informed collaboration.

\begin{table*}[t]
\centering

\begin{minipage}[t]{0.48\textwidth}
\centering
\captionof{table}{Long-term Task Utility}
\label{tab:task_utility_lite}
\small
\resizebox{\linewidth}{!}{
\begin{tabular}{llccc}
\toprule
Scenario & Protocol & Task Score & Completion Rate (\%) & Tokens/task (k) \\
\midrule
\multirow{3}{*}{Coding}
& A2A       & 26.03 & 53.12 & 4.5 \\
& Broadcast & 41.47 & 68.75 & 9.8 \\
& AEP       & 97.97 & 97.86 & 5.6 \\
\midrule
\multirow{3}{*}{Daily-Service}
& A2A       & 30.78 & 40.62 & 4.3 \\
& Broadcast & 38.28 & 59.38 & 7.2 \\
& AEP       & 93.19 & 90.62 & 7.4 \\
\midrule
\multirow{3}{*}{Research}
& A2A       & 35.38 & 71.88 & 3.4 \\
& Broadcast & 47.22 & 75.00 & 10.4 \\
& AEP       & 94.19 & 93.75 & 6.6 \\
\bottomrule
\end{tabular}
}
\end{minipage}
\hfill
\begin{minipage}[t]{0.48\textwidth}
\centering
\captionof{table}{Robustness under Malicious Agents}
\label{tab:scalable_attack_robustness}
\small
\resizebox{\linewidth}{!}{
\begin{tabular}{lccc}
\toprule
Protocol & Malicious Ratio $\rho$ & $\mathrm{nTSD}$ $\downarrow$
& $\mathrm{MCR}_{\mathrm{early}} \rightarrow \mathrm{MCR}_{\mathrm{late}}$ $\downarrow$ \\
\midrule
A2A       & 0.1 & 0.017 & $0.333 \rightarrow 0.333$ \\
A2A + Guard & 0.1 & 0.357 & $0.333 \rightarrow 0.333$ \\
A2A       & 0.4 & 1.000 & $0.333 \rightarrow 0.333$ \\
A2A + Guard & 0.4 & 0.500 & $0.333 \rightarrow 0.333$ \\
Broadcast & 0.1 & 0.000 & $0.133 \rightarrow 0.133$ \\
Broadcast + Guard & 0.1 & 0.000 & $0.133 \rightarrow 0.133$ \\
Broadcast & 0.4 & 0.350 & $0.400 \rightarrow 0.400$ \\
Broadcast + Guard & 0.4 & 0.300 & $0.400 \rightarrow 0.400$ \\
AEP       & 0.1 & \textbf{0.000} & $\mathbf{0.167 \rightarrow 0.000}$ \\
AEP       & 0.4 & \textbf{0.035} & $\mathbf{0.500 \rightarrow 0.000}$ \\
\bottomrule
\end{tabular}
}
\end{minipage}
\vskip -1.5em
\end{table*}

\subsection{Robustness under Increasing Malicious-Agent Ratios}
\label{sec:robustness_scalable_attacks}

We evaluate robustness on middle-scale Coding with malicious-agent ratios $\rho\in\{0,0.1,0.4\}$, where $\rho=0$ is the clean reference. Malicious agents behave normally during discovery but return benign-looking corrupted payloads once contacted, following prior poisoning attacks~\citep{amayuelas2024multiagent,cui2025mad,dong2025memory,chen2024agentpoison}. We compare A2A and Broadcast with external Qwen3Guard-Gen-0.6B defense, while AEP uses only its native controls, and report normalized Task Score Drop (nTSD) and early-to-late Malicious Contact Rate (MCR). Attack construction and metric are detailed in Appendix~\ref{app:robustness_scalable_attacks} and~\ref{app:metric-definitions}.
Table~\ref{tab:scalable_attack_robustness} shows that AEP remains robust as $\rho$ increases. At $\rho=0.4$, A2A and Broadcast incur nTSDs of $1.000$ and $0.350$, while AEP limits the drop to $0.035$. External guarding does not reduce malicious contact over time, whereas AEP drives $\mathrm{MCR}_{\mathrm{late}}$ to zero at both attack ratios. This suggests that robustness requires progressively identifying and avoiding unreliable collaborators, not only filtering individual messages.

\subsection{Ablation Study}

\begin{table*}[t]
\centering
\small
\caption{Module-level ablation results of AEP.}
\label{tab:aep_ablation_micro}

\begin{minipage}[t]{0.48\textwidth}
\centering
{\scriptsize (a) Asset Exchange ablation }

\vspace{0.4em}
\resizebox{\linewidth}{!}{
\begin{tabular}{lccc}
\toprule
Variant 
& Task Score $\uparrow$ 
& Completion (\%) $\uparrow$ 
& Tokens/task (k) $\downarrow$ \\
\midrule
AEP w/o Exchange 
& 98.3 
& 100.0 
& 4.5 \\
Full AEP 
& 97.9 
& 100.0 
& 3.5 \\
\midrule
$\Delta$ 
& $-0.4\%$ 
& $0.0\%$ 
& $-22.2\%$\\
\bottomrule
\end{tabular}
}
\end{minipage}
\hfill
\begin{minipage}[t]{0.48\textwidth}
\centering
{\scriptsize (b) Safety Gate ablation }

\vspace{0.4em}
{
\resizebox{\linewidth}{!}{
\setlength{\tabcolsep}{14pt}
\begin{tabular}{lcc}
\toprule
Variant 
& Privacy leakage $\downarrow$ 
& Injection success $\downarrow$ \\
\midrule
AEP w/o Safety Gate 
& 35.6\% 
& 100.0\% \\
Full AEP 
& 0.0\% 
& 0.0\% \\
\midrule
$\Delta$ 
& $-35.6\%$ 
& $-100.0\%$ \\
\bottomrule
\end{tabular}
}
}
\end{minipage}

\vspace{0.9em}

\begin{minipage}[t]{0.48\textwidth}
\centering
{\scriptsize (c) Routing ablation }

\vspace{0.4em}
\resizebox{\linewidth}{!}{
\begin{tabular}{lccc}
\toprule
Variant 
& Task Score $\uparrow$ 
& Completion (\%) $\uparrow$ 
& Tokens/task (k) $\downarrow$ \\
\midrule
AEP w/o Routing
& 27.78
& 22.22
& 5.6 \\
Full AEP
& 97.44
& 100.00
& 3.5 \\
\midrule
$\Delta$
& +250.8\%
& +350.0\%
& -37.5\% \\
\bottomrule
\end{tabular}
}
\end{minipage}
\hfill
\begin{minipage}[t]{0.48\textwidth}
\centering
{\scriptsize (d) Coordination ablation}

\vspace{0.4em}
\resizebox{\linewidth}{!}{
\begin{tabular}{lccc}
\toprule
Variant 
& Task Score $\uparrow$ 
& Completion (\%) $\uparrow$ 
& Tokens/task (k) $\downarrow$ \\
\midrule
AEP w/o Coordination
& 33.33
& 100.00
& 2.1 \\
Full AEP
& 97.44
& 100.00
& 3.5 \\
\midrule
$\Delta$
& +192.3\%
& 0.0\%
& +66.7\% \\
\bottomrule
\end{tabular}
}
\end{minipage}

\end{table*}

Table~\ref{tab:aep_ablation_micro} reports ablations of AEP on the middle-scale network. Removing Asset Exchange preserves task performance and 100\% completion, but increases token cost from 3.5k to 4.5k per task, indicating that it mainly improves communication efficiency. Removing the Safety Gate causes privacy leakage to rise from 0.0\% to 35.6\% and injection success from 0.0\% to 100.0\%, confirming its critical role in security. Routing and Coordination are important for task quality: without Routing, the Task Score drops to 27.78 and Completion Rate to 22.22\%, while removing Coordination lowers the Task Score to 33.33 despite retaining 100\% completion. We do not ablate AEP-Ledger because it serves as the protocol substrate, providing the logical clock and credit validation. Removing it would invalidate AEP’s execution mechanisms rather than isolate a single module’s contribution.

\section{Related Work}

\noindent\textbf{LLM-based Multi-Agent Systems.}
Existing LLM-based multi-agent systems explore conversational, role-based, workflow-driven, distributed, and debate-based collaboration~\citep{wu2024autogen,li2023camel,chen2024agentverse,hong2023metagpt,qian2024chatdev,gao2024agentscope,du2024improving,kaesberg2025voting}.
Most assume known participants, communication patterns, or shared orchestration environments, whereas open private agent networks require collaborator discovery, reachability and trust reasoning, and coordination under partial topology and ownership boundaries.

\noindent\textbf{Private Agent Cooperation.}
Recent work has studied heterogeneous agent ecosystems, agent communication protocols, personal agents, decentralized LLM-based multi-agent systems, and controlled memory sharing~\citep{chen2024internet,yang2025survey,li2024personal,yang2025unlocking,rezazadeh2025collaborative}.
Request-centric protocols, such as agent-to-agent invocation through known endpoints or agent cards~\citep{a2aprotocol2025, chang2025agent, yuan2026beyond}, support capability advertisement and task execution, but usually assume that the requester already knows whom to contact.
Shared-memory, broadcast, blackboard, and tightly coordinated systems~\citep{eigenflux2026whitepaper,moltbook2026,han2025exploring,masterman2024landscape} enable information reuse, but rely on stronger assumptions about shared state, topology visibility, governance, and trusted infrastructure. These assumptions are difficult to satisfy in open private agent networks, where agents must collaborate across ownership boundaries under partial observability.

\noindent\textbf{Safety Risks in Multi-Agent Collaboration.}
Agentic systems face multiple risks, including privacy leakage~\citep{yagoubi2026agentleak,wang2025privacy,zhang2025searching,wang2025silent,wang2026mempot}, prompt injection~\citep{zhan2024injecagent,debenedetti2024agentdojo}, memory poisoning~\citep{dong2025memory,chen2024agentpoison}, and adversarial multi-agent collaboration attack~\citep{amayuelas2024multiagent,cui2025mad}.
These risks can propagate through inter-agent messages, shared memory, delegated tool use, and long-term memory updates.
Open private agent networks therefore require cooperation mechanisms that discover useful and trustworthy collaborators under partial observability, while controlling both what private agents release and what they accept from others.

\section{Conclusion}

We introduced OPAC, an asset-centric paradigm for cooperation among private agents in open, partially observed, and low-trust networks, and instantiated it with AEP, which combines semantic discovery, demand-driven coordination, safety-gated asset exchange, and ledger-backed evidence derivation.
Experiments show that AEP improves collaboration utility while reducing privacy leakage and injection risk, providing a practical foundation for trustworthy agent-native communication and networking.

\section*{Acknowledgments}
This work was conducted solely for academic research purposes, and no harm or disruption was caused to real-world networks or systems during the experiments.

\noindent We thank Ying Ling from Southern University of Science and Technology for her insightful comments and valuable assistance with the experimental evaluation.

\clearpage
\appendix

\section*{Appendix Contents}
\addcontentsline{toc}{section}{Appendix Contents}

\startcontents[appendix]

\printcontents[appendix]{}{1}{\setcounter{tocdepth}{3}}

\clearpage

\section{Extended Operational View of \textsc{AEP}}
\label{app:aep-operational}

\begin{figure}[t]
    \centering
    \includegraphics[width=1\linewidth]{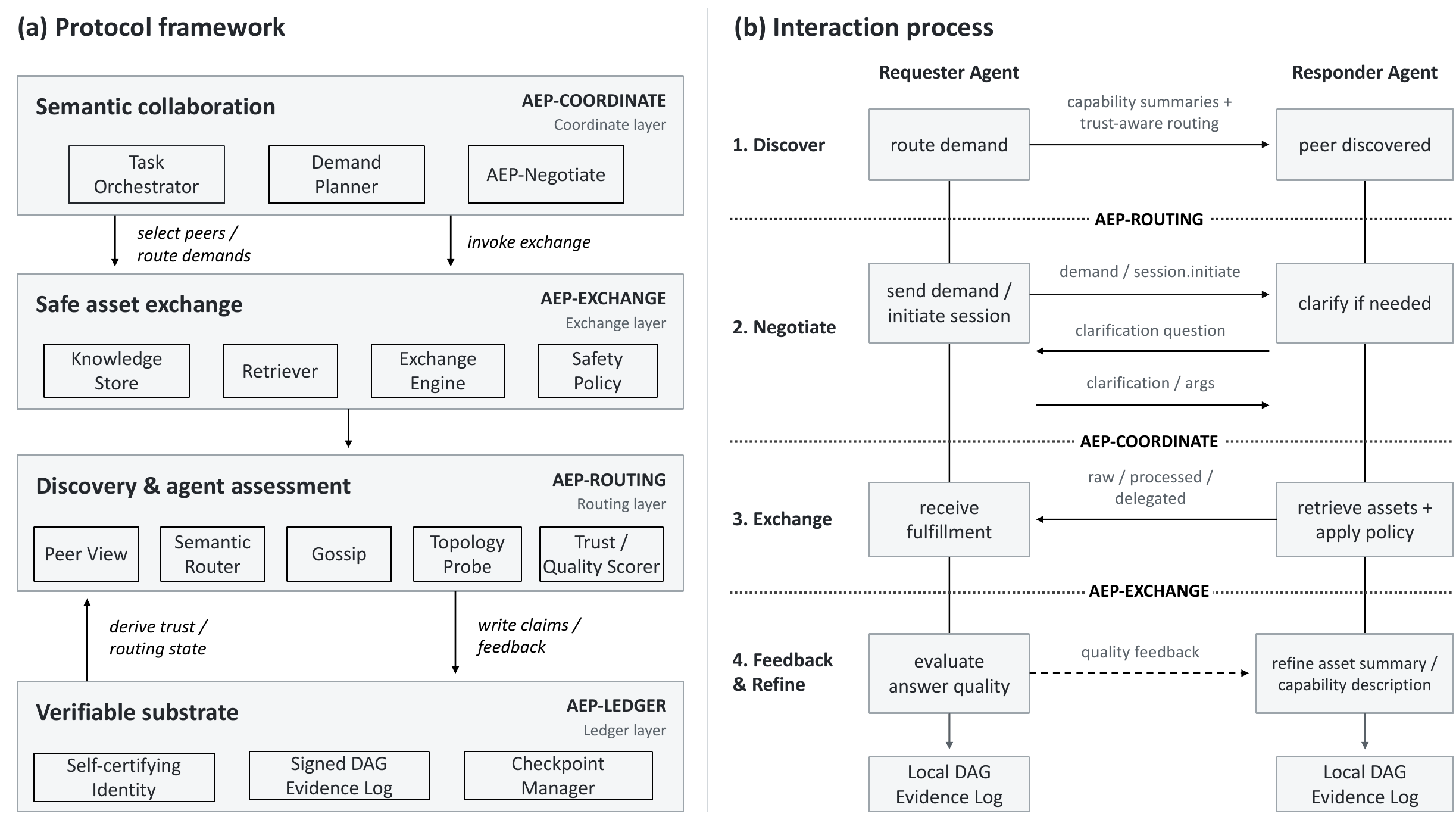}
    \caption{Protocol Framework and Sequence Diagram in Interaction Process.}
    \label{fig:framework}
\end{figure}

This appendix gives an implementation-oriented view of Section~3 while preserving the higher-level abstractions used in the main text.
Table~\ref{tab:aep-instantiation} summarizes how the abstractions used in Section~3 map onto concrete implementation components in the current protocol stack. Figure~\ref{fig:framework} represents the full structure of our protocol and the interaction process with all protocol modules.

\begin{table}[t]
\centering
\caption{Implementation-oriented instantiation of the abstractions used in Section~3. The table shows how protocol-level concepts correspond to concrete runtime components without elevating those runtime details to the main theoretical contribution.}
\footnotesize
\setlength{\tabcolsep}{4pt}
\begin{tabular}{@{}p{0.34\linewidth}p{0.58\linewidth}@{}}
\toprule
Paper abstraction & Implementation component \\
\midrule
Cognitive map / semantic flow table & \texttt{RoutingProtocol} + \texttt{FlowTable} \\
Evidence-derived candidate admission & Derived verified routing view + local trust / quality / blacklist state \\
Demand planning and allocation & \texttt{TaskOrchestrator} + demand backlog + adaptive Top-$K$ policy \\
Negotiated exchange & \texttt{AepNegotiateSession} \\
Asset exchange & \texttt{ExchangeEngine} + \texttt{SafetyBackend} \\
Asynchronous cooperation & \texttt{SemanticJobService} + one-way fulfill \\
Audited evidence & Signed hash-linked DAG / Merkle DAG + evidence verification pipeline \\
Trust coordinate $T(\pi)$ & Evidence-derived local safety trust over verified interaction evidence \\
Quality feedback & Quality-feedback records + derived quality score + summary / capability refinement \\
Evidence-log compression & Quorum-backed checkpoint commit + DAG pruning \\
\bottomrule
\end{tabular}
\label{tab:aep-instantiation}
\end{table}

\subsection{\textsc{AEP-Routing} Operational Details}
\label{app:aep-routing}

\subsubsection{Semantic Flow Table}
\label{app:semantic-flow-table}
The Semantic Flow Table is the local routing state used by AEP-Routing to support
semantic peer discovery, safety-aware candidate admission, and quality-aware reranking.
Each entry corresponds to a discovered peer agent and stores compact public metadata rather
than private memory.  The table combines a raw routing view, represented by
\texttt{FlowEntry}, with an optional ledger-derived verified view, represented by
\texttt{DerivedFlowEntry}.  The raw view records information learned from gossip,
bootstrap configuration, query responses, or local announcements, including agent identity,
capability description, endpoint information, route freshness, route confidence, public
memory summary, keywords, and skill summary.  The derived view is reconstructed from the
local evidence ledger and overlays the raw routing state with safety-relevant evidence,
including blacklist status, verified connectivity, evidence-derived trust, service quality,
and provenance.  During routing, blacklist status and trust determine whether a peer is
admissible, while semantic similarity and service quality determine the ranking among
already-admissible peers.  This separation ensures that a useful but unsafe peer cannot be
promoted solely by utility, while a safe but low-quality peer is not incorrectly treated as
malicious.

\begin{table}[t]
\centering
\scriptsize
\setlength{\tabcolsep}{4pt}
\renewcommand{\arraystretch}{1.12}
\begin{tabularx}{\textwidth}{
    >{\ttfamily\raggedright\arraybackslash}p{0.22\textwidth}
    >{\raggedright\arraybackslash}p{0.22\textwidth}
    X}
\toprule
\normalfont Field & View & Description \\
\midrule
agent\_id &
Raw / Derived &
Globally unique identifier of the peer agent, typically tied to the peer's
self-certifying public identity. \\

capability\_desc &
Raw / Derived &
Public capability description announced by the peer. This is the primary textual signal
used for semantic routing. \\

trust\_score &
Raw / Derived &
Local or evidence-derived trust estimate. In routing, the derived trust score is preferred
when available and is used for candidate admission. \\

endpoint\_url &
Raw / Derived &
Network endpoint of the peer, such as an HTTPS JSON-RPC address. In the derived view,
this endpoint is extracted from the latest verified announcement. \\

next\_hop &
Raw &
Next-hop agent identifier used to reach the peer. In the current direct-exchange setting,
this is usually equal to \texttt{agent\_id}. \\

hops &
Raw &
Estimated hop distance from the current node to the peer. It is also used as a simple
route-cost proxy. \\

last\_seen\_clock &
Raw / Derived &
Logical clock of the most recent observation or verified announcement. It is used for
freshness control and stale-entry decay. \\

agent\_card &
Raw &
Optional full Agent Card object, preserving extensible metadata not normalized into the
core routing fields. \\

learned\_via &
Raw &
Source from which the entry was learned, such as \texttt{gossip}, \texttt{bootstrap},
\texttt{query\_response}, \texttt{restored}, or \texttt{local}. \\

announce\_seq &
Raw &
Sequence number of the corresponding \texttt{capability\_announce} block, used for
freshness comparison and conflict arbitration. \\

path\_cost &
Raw &
Estimated routing cost, currently approximated by \texttt{float(hops)}. \\

route\_confidence &
Raw &
Confidence in the route entry. Gossip-propagated entries may decay with hop distance,
for example using a rule such as \texttt{max(0.2, 1/(1+hops))}. \\

memory\_summary &
Raw / Derived &
Public summary of the peer's memory state. It provides a secondary semantic matching
signal without exposing private memory. \\

keywords &
Raw / Derived &
Compact keyword tuple extracted from public capability or memory metadata, used for
lexical-overlap matching. \\

skill\_summary &
Raw / Derived &
Public summary of the peer's available skills, such as a skill-name list or capability
signature. \\

skill\_count &
Raw / Derived &
Number of skills declared by the peer, used as an auxiliary capability-richness signal. \\

source\_block\_hash &
Raw / Derived &
Hash of the \texttt{capability\_announce} evidence block from which the entry was learned
or derived. It supports provenance tracing and endpoint verification. \\

is\_connected &
Derived &
Whether the peer has evidence-backed reachability, for example through matched topology
probe evidence or matched exchange claims. \\

is\_blacklisted &
Derived &
Whether the peer is locally blacklisted due to equivocation, double signing, safety
incidents, or other ledgered evidence. \\

service\_quality &
Derived &
Quality score smoothed from historical feedback. It is used mainly to rerank
already-admissible peers, not to override safety trust. \\
\bottomrule
\end{tabularx}
\vspace{0.5em}
\caption{Fields maintained by the Semantic Flow Table. The raw view supports routing and
semantic matching, while the derived view adds ledger-backed safety and quality signals.}
\label{tab:semantic-flow-table-fields}
\end{table}

\subsubsection{Constrained Gossip Synchronization}
\label{app:gossip-sync}

This subsection describes the operational gossip procedure used by
\textsc{AEP-Routing}. Gossip is responsible for disseminating both lightweight
routing metadata and verifiable evidence state under partial topology. Its design
decouples fast semantic propagation from slower evidence verification, while
bounding communication amplification.

\noindent\textbf{Gossip schedule.}
Each agent periodically runs a gossip step as part of its routing runtime. At
the beginning of a gossip cycle, the agent advances its logical clock, removes
stale routing entries, recomputes evidence-derived trust if needed, selects a
small set of gossip peers, constructs synchronization envelopes, and enqueues
them for outbound transmission. Staleness pruning prevents old flow entries from
being repeatedly propagated, while periodic trust recomputation incorporates new
ledger evidence into later peer selection.

\noindent\textbf{Peer selection.}
Let $\mathcal{N}_i(t)$ be the set of peers currently visible to agent $A_i$.
AEP selects a bounded fanout set by trust-weighted sampling without replacement:
\begin{equation}
    p_{ij}(t)
    \propto
    \max\{\epsilon, \mathrm{Trust}_{ij}(t)\},
    \qquad A_j \in \mathcal{N}_i(t),
\end{equation}
where $\epsilon>0$ ensures that low-trust but not-yet-excluded peers are not
assigned exactly zero sampling probability. The fanout budget limits per-round
amplification. To reduce eclipse risk, the implementation also tracks recent
gossip sources; if source diversity falls below a minimum threshold, the sender
temporarily expands the peer set with additional randomly selected peers outside
the initial sample.

\noindent\textbf{Synchronization envelope.}
A gossip message is a synchronization envelope with three payload components:
\begin{equation}
    M_i =
    \bigl(
    \texttt{entries},
    \texttt{blocks},
    \texttt{heads}
    \bigr).
\end{equation}
The \texttt{entries} field contains recently active \texttt{FlowEntry} records
exported from the sender's raw semantic flow table. Only entries whose
last-observed clock is within a freshness window are exported:
\begin{equation}
    \ell_i(t) - \ell^{\mathrm{seen}}_j \leq s_{\mathrm{export}}.
\end{equation}
The \texttt{blocks} field contains signed DAG evidence blocks that the receiver
is estimated not to have yet. These blocks are exported incrementally using the
receiver's previously known DAG heads. Blocks are sorted by logical order and
included greedily subject to a maximum block count and maximum serialized payload
size. The optional \texttt{heads} field contains the sender's current DAG heads,
which allow the receiver to update its view of the sender's ledger frontier and
enable future delta synchronization.

\noindent\textbf{Inbound processing.}
Upon receiving a gossip envelope, the receiver first performs cheap defensive
checks. It discards duplicate messages using a bounded message-id cache, rejects
messages exceeding block-count or payload-size limits, and records the sender in
the local gossip-source tracker.
The receiver then processes the envelope along two paths. The fast path merges
\texttt{FlowEntry} records into the raw semantic flow table. During this merge,
the receiver rewrites the next hop of each imported route to the immediate
sender, increments the hop count, and updates route confidence as a decreasing
function of hop length and evidence availability. If multiple entries describe
the same target agent, the receiver arbitrates among them using freshness,
announcement sequence, and path length.
The slow path verifies and ingests signed DAG blocks. For each received block,
the receiver checks the author signature, validates any announcement-specific
proof requirements, inserts the block into the local evidence DAG, and applies
accepted side effects. For example, trust-update blocks may update distributed
safety assessments, while quality-feedback blocks may update local service
quality estimates. If a flow entry points to a source block that has not yet
arrived, the entry remains only an unverified routing cache entry until the
corresponding evidence is available.

\noindent\textbf{Head synchronization.}
When a gossip envelope contains DAG heads, the receiver accepts them only if the
heads are locally known or already covered by a finalized checkpoint. Accepted
heads update the receiver's view of the sender's ledger frontier. Future gossip
messages can then export only blocks after that frontier, avoiding repeated
transmission of already synchronized evidence.

\noindent\textbf{Burst propagation.}
If merging incoming flow entries changes the local routing table, the agent may
temporarily enter a short burst window. During this window, it gossips more
aggressively than the normal periodic schedule, allowing newly discovered routing
state to propagate quickly. The burst is bounded by a small logical-clock budget,
so it accelerates convergence without turning gossip into unbounded broadcast.

\noindent\textbf{Summary.}
The gossip mechanism therefore separates semantic and evidential synchronization.
Flow entries propagate quickly and support early local routing views, while DAG
blocks provide delayed but verifiable evidence for authentication, trust,
quality, and topology inference. Fanout control, freshness windows, TTL,
payload caps, deduplication, source-diversity tracking, and bounded burst
broadcast jointly keep synchronization scalable and robust under partial
topology.

\subsubsection{Topology Probe}
\label{app:topology-probe}

Topology probing is the active reachability-verification mechanism used by
\textsc{AEP-Routing}. Its purpose is to distinguish a peer that is merely learned
through gossip from a peer that can support direct point-to-point semantic
exchange. A successful probe produces signed evidence that can later be consumed
by routing, exchange, and trust inference.

\noindent\textbf{Trigger condition.}
During local view synchronization, agent $A_i$ considers probing a discovered
peer $A_j$ when $A_j$ has a non-empty endpoint, is not already connected in the
derived evidence view, has not already been confirmed by a previous probe, and
has not been probed for the same announcement version. The announcement version
can be represented by a probe key such as
\begin{equation}
    \mathrm{key}_{ij}
    =
    (A_j,\mathrm{source\_block\_hash}_j,\mathrm{announce\_seq}_j).
\end{equation}
This prevents repeated probes against the same stale routing announcement while
allowing a new probe when the peer publishes a fresh capability announcement.

\noindent\textbf{Signed probe handshake.}
A probe is a lightweight signed request--response exchange. The requester sends
a message containing a probe identifier, source and target identifiers, a nonce,
the requester's latest capability-announcement block hash, and a logical clock.
The probe identifier can be written abstractly as
\begin{equation}
    \mathrm{probe\_id}
    =
    H(A_i \,\|\, A_j \,\|\, h_i^{\mathrm{cap}} \,\|\, \ell_i),
\end{equation}
where $h_i^{\mathrm{cap}}$ is the requester's latest capability block hash and
$\ell_i$ is its logical clock. The request is signed by $A_i$.

Upon receipt, $A_j$ checks that the target identity matches itself, verifies that
the claimed source identity matches the transport peer, validates the request
signature using the source public key, and rejects duplicate probe identifiers.
If the request is accepted, $A_j$ returns a signed response echoing the
\texttt{probe\_id} and nonce, adding a fresh response nonce, its own latest
capability-announcement block hash, and an acceptance status.

\noindent\textbf{Evidence recording.}
Both sides record the successful handshake as ledger evidence. Let
\begin{equation}
    h_{ij}^{\mathrm{probe}}
    =
    H(\mathrm{probe\_id}, A_i, A_j, \mathrm{nonce})
\end{equation}
be the shared probe hash. The receiver records a block of kind
\texttt{topology\_probe\_rx}, while the requester records a block of kind
\texttt{topology\_probe\_tx}. These records have the same logical role as
ordinary exchange claims: only matched transmit and receive claims for the same
hash support an edge in the evidence-backed topology.

\noindent\textbf{Topology effect.}
The topology inference procedure treats matched topology-probe claims as
evidence of a verified undirected edge:
\begin{equation}
    (A_i,A_j)\in G_i^{\mathrm{ev}}(t)
    \quad\Longleftarrow\quad
    \exists h:
    \mathrm{tx}(A_i,A_j,h)\wedge \mathrm{rx}(A_j,A_i,h).
\end{equation}
Thus, a successful topology probe can promote a gossip-discovered peer into an
evidence-reachable peer without requiring a prior semantic asset exchange.

\noindent\textbf{Use in routing and exchange.}
A probe-confirmed peer is treated as directly reachable for collaboration
eligibility, even if it was originally learned through a multi-hop gossip path.
The confirmation is also exposed to exchange-side evidence construction, where
probe confirmation and derived connectivity are used as alternative forms of
topology confirmation. After probe evidence is recorded, trust inference may be
recomputed because $G_i^{\mathrm{ev}}(t)$ has gained a new verified edge.

\noindent\textbf{Security role.}
Topology probes do not transfer task data and do not update service quality.
Their role is narrower: they authenticate the peer identity, test direct
network reachability, prevent replay through nonces and processed-probe caches,
and create verifiable evidence for later topology inference. This keeps routing
from relying on unverified multi-hop gossip alone while avoiding the cost of
broadcasting semantic requests to unconfirmed peers.

\subsubsection{Remote Agent Selection}
\label{app:remote-agent-selection}

This subsection specifies the remote-agent selection procedure used by
AEP-ROUTING. Given an unresolved demand from requester $A_i$, the selector
returns a small set of peers that are semantically relevant, safety-admissible,
and practically reachable. The selector only decides whom to contact; whether
any asset is released is still decided later by AEP-EXCHANGE.

Let $q_i$ be the embedding of the current demand. Agent $A_i$ maintains a raw
semantic flow table $\mathcal{T}_i$ and, when ledger evidence is available, a
derived verified view $\mathcal{D}_i$. For each candidate $A_j$, the raw table
provides public semantic and routing metadata, including capability summary
$\mathrm{cap}_j$, memory summary $\mathrm{mem}_j$, skill summary
$\mathrm{skill}_j$, keywords $\mathrm{kw}_j$, endpoint, hop count, next hop,
freshness, and raw trust. The derived view adds evidence-derived trust
$\tilde{\tau}_{ij}$, service quality $\tilde{q}_{ij}$, blacklist status, and
verified connectivity.

\noindent\textbf{Candidate admission.}
Selection first applies hard safety gates. A peer is rejected if it is
blacklisted, lacks a usable endpoint, or fails the effective trust threshold.
The effective trust score is
\begin{equation}
\tau^{\mathrm{eff}}_{ij} =
\begin{cases}
\tilde{\tau}_{ij}, & \text{if verified connectivity is available},\\
0.5\tilde{\tau}_{ij}, & \text{if derived evidence exists but connectivity is unverified},\\
0.3\tau_{ij}, & \text{if only raw routing evidence is available}.
\end{cases}
\end{equation}
The admitted candidate set is
\begin{equation}
\mathcal{C}_i =
\left\{
A_j \in \mathcal{T}_i :
A_j \notin \mathrm{Blacklist}_i,\;
\tau^{\mathrm{eff}}_{ij} \geq \tau_{\min}
\right\}.
\end{equation}
Thus, trust is used as an admission signal rather than only as a ranking feature.

\noindent\textbf{Semantic scoring.}
For every admitted peer, AEP computes a semantic relevance score from public
metadata:
\begin{equation}
    s_{ij}
    =
    w_c \textrm{sim}(q_i,\mathrm{cap}_j)
    +
    w_m \textrm{sim}(q_i,\mathrm{mem}_j)
    +
    w_s \textrm{sim}(q_i,\mathrm{skill}_j)
    +
    w_k\,\mathrm{Jaccard}(K_i,\mathrm{kw}_j),
\end{equation}

where $K_i$ is the keyword set extracted from the demand, $\textrm{sim}(\cdot)$ is the cosine similarity with sentence embedding. Missing metadata fields
are ignored and the remaining weights are renormalized. If service-quality
evidence is available, the semantic score is adjusted as
\begin{equation}
\bar{s}_{ij}
=
s_{ij}\bigl(0.7 + 0.3\tilde{q}_{ij}\bigr).
\end{equation}
Service quality therefore reranks already-admitted peers, but cannot rescue a
blacklisted or insufficiently trusted peer.

\noindent\textbf{Topology filtering and Top-$k$ selection.}
After scoring, the selector keeps only candidates that are currently usable for
remote collaboration. A peer is route-eligible if it is directly reachable,
confirmed by topology-probe evidence, or connected through an admissible route
within the collaboration hop limit $h_{\max}$\footnote{In this version, we assume all agents with authorized collaboration are connected and agents are not allowed to collaborate if they are not mutually reachable, therefore $h_{\max}$ is default set to $1$.}. The final remote candidate set is
\begin{equation}
\mathcal{R}_i =
\left\{
A_j \in \mathcal{C}_i :
\mathrm{RouteEligible}_i(A_j; h_{\max}) = 1
\right\}.
\end{equation}
AEP-ROUTING returns
\begin{equation}
\mathrm{Select}_i(q_i)
=
\operatorname{TopK}_{A_j \in \mathcal{R}_i}
\bar{s}_{ij}.
\end{equation}

\begin{algorithm}[t]
\caption{Remote Agent Selection in AEP-ROUTING}
\label{alg:remote-agent-selection}
\small
\textbf{Require:} demand embedding $q_i$, raw table $\mathcal{T}_i$, derived
view $\mathcal{D}_i$, trust floor $\tau_{\min}$, hop limit $h_{\max}$, budget $k$
\begin{tabbing}
00\quad \= 00\quad \= 00\quad \= \kill
1: \> $\mathcal{C}_i \gets \emptyset$\\
2: \> \textbf{for} $A_j \in \mathcal{T}_i$ \textbf{do}\\
3: \> \> \textbf{if} $A_j$ is blacklisted or has no usable endpoint \textbf{then}\\
4: \> \> \> continue\\
5: \> \> \textbf{end if}\\
6: \> \> compute effective trust $\tau^{\mathrm{eff}}_{ij}$\\
7: \> \> \textbf{if} $\tau^{\mathrm{eff}}_{ij} \geq \tau_{\min}$ \textbf{then}\\
8: \> \> \> add $A_j$ to $\mathcal{C}_i$\\
9: \> \> \textbf{end if}\\
10: \> \textbf{end for}\\
11: \> \textbf{for} $A_j \in \mathcal{C}_i$ \textbf{do}\\
12: \> \> compute semantic score $s_{ij}$\\
13: \> \> adjust by service quality to obtain $\bar{s}_{ij}$\\
14: \> \textbf{end for}\\
15: \> remove peers that fail $\mathrm{RouteEligible}_i(A_j;h_{\max})$\\
16: \> \textbf{return} top-$k$ remaining peers ranked by $\bar{s}_{ij}$
\end{tabbing}
\end{algorithm}
This procedure implements the separation used by AEP-Routing: semantic metadata
expands discovery, trust determines candidate admissibility, quality ranks
safe candidates, and topology evidence prevents unreachable or weakly supported
routes from entering downstream coordination.

\subsubsection{Evidence-Derived Peer Rating and Pruning}
\label{app:evidence-derived-rating-pruning}

This subsection describes the operational procedure behind evidence-derived
peer rating and pruning in \textsc{AEP-Routing}. The procedure consumes the raw
semantic flow table $\mathcal{T}_i(t)$, the local evidence view
$\mathcal{Q}_i(t)$, and the evidence-backed interaction topology
$G_i^{\mathrm{ev}}(t)$.

\noindent\textbf{Evidence-backed topology.}
Agent $A_i$ distinguishes announced reachability from verified reachability.
The evidence topology $G_i^{\mathrm{ev}}(t)$ is constructed only from matched
two-sided evidence, such as paired exchange claims or paired topology-probe
claims. Thus, a peer is treated as evidence-reachable only when the ledger
contains verifiable support for the corresponding interaction edge.

\noindent\textbf{Safety trust.}
After audited interactions, $A_i$ records bounded local safety assessments for
its peers. Remote trust reports may also be ingested when they appear in valid
signed evidence blocks. These local and distributed assessments are then
aggregated over $G_i^{\mathrm{ev}}(t)$ by a damped propagation procedure:
\begin{equation}
p_i^{(\ell+1)}
=
(1-\alpha)\frac{\mathbf{1}}{n_i}
+
\alpha P_i(t)^\top p_i^{(\ell)},
\end{equation}
where $P_i(t)$ is a transition matrix induced by evidence-backed safety
assessments, which is usually records normalized peer rating value coming from disseminated ledger evidence. After convergence, the derived trust score of peer $A_j$ is
\begin{equation}
\mathrm{Trust}_{ij}(t)=p^\star_{i,j}.
\end{equation}
Thus, trust is a topology-derived safety rating rather than a raw average of
recent interactions.

\noindent\textbf{Blacklist state.}
Blacklist status is derived from verifiable protocol violations. For example,
if the evidence log contains equivocation evidence, such as conflicting signed
blocks from the same author at the same sequence position, the corresponding
peer is marked as blacklisted and removed from future candidate sets.

\noindent\textbf{Pruning.}
Pruning is implemented as consecutive filters over $\mathcal{T}_i(t)$.

First, stale entries are removed:
\begin{equation}
\mathcal{T}^{\mathrm{fresh}}_i(t)
=
\left\{
A_j\in\mathcal{T}_i(t):
\ell_i(t)-\ell^{\mathrm{seen}}_j \le s_{\max}
\right\}.
\end{equation}

Second, unsafe entries are excluded using blacklist status and an effective trust
score:
\begin{equation}
\mathcal{T}^{\mathrm{safe}}_i(t)
=
\left\{
A_j\in\mathcal{T}^{\mathrm{fresh}}_i(t):
A_j\notin\mathrm{Blacklist}_i(t),\;
\tau^{\mathrm{eff}}_{ij}(t)\ge\tau_{\min}
\right\}.
\end{equation}
Here $\tau^{\mathrm{eff}}_{ij}(t)$ is derived from verified trust when available,
discounted when evidence is incomplete, and falls back to raw routing trust only
when no verified view exists.

Third, unreachable entries are removed:
\begin{equation}
\mathcal{P}_i(t)
=
\left\{
A_j\in\mathcal{T}^{\mathrm{safe}}_i(t):
\mathrm{Reachable}_i(A_j)=1
\right\}.
\end{equation}
A peer is reachable if it has a usable endpoint and is directly connected,
bootstrap-connected, topology-probe confirmed, or connected in
$G_i^{\mathrm{ev}}(t)$.

The output $\mathcal{P}_i(t)$ is therefore a fresh, non-blacklisted,
sufficiently trusted, and evidence-reachable peer set. Later semantic routing
can rank this set for a specific task, but the rating and pruning procedure
itself is evidence-driven and task-independent.

\subsection{\textsc{AEP-Coordinate} Operational Details}
\label{app:aep-coordinate}

This appendix specifies the operational procedure behind
\textsc{AEP-Coordinate}. The coordinator controls when collaboration is needed,
how unresolved needs are represented, how peers are allocated, and how feedback
is converted into later refinement.

\noindent\textbf{Local sufficiency.}
When a task query $x_i$ arrives, requester $A_i$ first retrieves local memory and
skill context. The coordinator estimates whether this local context is sufficient
for answering the task. In the full implementation, this can be decided by an
LLM gate that receives the task, local context, and retrieval similarity, and
returns a structured decision
\begin{equation}
    (\mathrm{sufficient},\mathrm{confidence},\mathrm{reason}).
\end{equation}
If this gate is unavailable or fails, the coordinator falls back to a similarity
threshold over the top retrieved local assets. If the task is locally sufficient,
the task is answered without entering the collaboration loop.

\noindent\textbf{Demand planning.}
If local context is insufficient, the coordinator converts the task into a demand
backlog
\begin{equation}
    \mathcal{B}_i=\{d_1,\ldots,d_m\}.
\end{equation}
Each demand $d$ contains a demand identifier, a textual description, a priority,
a status, and bookkeeping fields such as previously requested peers. A demand
may represent a missing fact, missing context fragment, executable subproblem,
or clarification need. The planner may condition on the original task, the
previous demand backlog, received responses, and the latest insufficiency reason.
If planning fails, the fallback is a single demand covering the original query.

\noindent\textbf{Candidate allocation.}
For each unresolved demand $d$, the coordinator requests a route pool from
\textsc{AEP-Routing}. The returned peers have already passed semantic and
topology-aware routing filters. The coordinator may optionally rerank the pool
using task-level evidence, then removes peers that were already tried for the
same demand:
\begin{equation}
    \mathcal{U}_{i,d}
    =
    \mathcal{C}_{i,d}
    \setminus
    \mathrm{RequestedPeers}_{i,d}.
\end{equation}
The remaining untried candidates are allocated according to the demand priority
and collaboration budget. If the caller provides preferred targets, the
coordinator bypasses semantic routing and only checks whether those targets are
eligible for direct collaboration. If routing returns no usable peers, the
implementation may fall back to the local agent view, ranked by effective trust,
freshness, and route confidence.

\noindent\textbf{Negotiation.}
If a demand is underspecified, the requester may open a short negotiation session
instead of immediately requesting an exchange. A session has an identifier,
a round counter, and a remaining budget. The responder may ask clarification
questions or request missing task parameters. The requester then decides whether
to continue negotiation, convert the session into a direct exchange request, or
terminate it. The loop is bounded by maximum rounds and budget:
\begin{equation}
    r \le r_{\max},\qquad b_r \ge 0.
\end{equation}
Negotiation therefore delays asset release until the requested context is better
specified, reducing unnecessary communication and accidental over-sharing. By default, the $r_{\max}$ and the $b_r$ is 3 and 9.

\noindent\textbf{Demand resolution.}
After receiving a response, the coordinator updates the demand status. A demand
is marked resolved when the response provides sufficient evidence with confidence
above a resolution threshold:
\begin{equation}
    \mathrm{Resolved}(d)=
    \mathbf{1}\{\mathrm{conf}(d)\ge \theta_{\mathrm{res}}\}.
\end{equation}
Otherwise, it remains pending, is retried with additional candidates, or becomes
exhausted if the candidate budget is depleted. By default, the sufficiency threshold is set to 0.75.

\noindent\textbf{Quality feedback and refinement.}
For each fulfilled demand, the requester computes a quality signal
$\Psi^{\mathrm{qual}}_{ij}\in[0,1]$ measuring the usefulness of peer $A_j$'s
response for the original task or demand. The feedback is used in two ways.
First, it updates the local quality estimate:
\begin{equation}
    \mathrm{Qual}_{ij}(t+1)
    =
    \mathrm{clip}_{[0,1]}
    \left(
    (1-\eta)\mathrm{Qual}_{ij}(t)
    +
    \eta\Psi^{\mathrm{qual}}_{ij}(t)
    \right).
\end{equation}
Second, it is recorded as ledger evidence so that other agents may ingest the
feedback and update their own local estimates. Repeated negative feedback can
trigger diagnosis of failure modes such as stale memory, capability mismatch,
or low-quality asset summaries. The implementation may then register asset
summaries or capability descriptions for refinement and publish improved
metadata in later capability announcements.

\noindent\textbf{Role in the protocol.}
\textsc{AEP-Coordinate} does not itself release assets. It determines whether
external cooperation is needed, decomposes the need into bounded demands,
allocates those demands to candidate peers, manages short negotiation loops, and
turns downstream outcomes into feedback for future routing and metadata
refinement.

\subsection{\textsc{AEP-Exchange} Operational Details}
\label{app:aep-exchange}

This appendix specifies the operational procedure behind
\textsc{AEP-Exchange}. The exchange layer is responsible for selecting relevant
assets, deciding the release mode, enforcing outbound and inbound safety gates,
and integrating admitted payloads into exchange memory.

\noindent\textbf{Candidate assets.}
For a demand from requester $A_i$, responder $A_j$ searches its local memory and
skills for candidate assets
\begin{equation}
    \Xi_j=\{\xi_1,\ldots,\xi_m\}.
\end{equation}
If no relevant asset is found, the exchange engine returns a no-relevant-asset
decision. Otherwise, candidates are pre-filtered by relevance and bounded by a
pre-evaluation budget so that expensive safety and LLM-based checks are applied
only to plausible assets.

\noindent\textbf{Trust and topology gate.}
Before evaluating individual assets, the responder constructs exchange evidence
for the requester. If the requester is blacklisted, the exchange is blocked. If
the requester lacks topology confirmation, its effective trust may be discounted.
The request is allowed to proceed only if
\begin{equation}
    \tau^{\mathrm{eff}}_{ji} \ge \tau_{\mathrm{floor}}.
\end{equation}
This gate prevents low-trust or unconfirmed peers from reaching the asset-release
stage.

\noindent\textbf{Relevance filtering and reranking.}
Each surviving candidate asset is scored against the demand. Assets whose
similarity falls below a minimum relevance threshold are removed. When enabled,
an additional LLM relevance check can reject superficially similar but
semantically mismatched assets. If many assets survive, the exchange engine may
rerank them with a task-aware reranker before safety evaluation.

\noindent\textbf{Release modes.}
For each candidate asset $\xi$, the safety backend returns a release decision:
\begin{equation}
    a(\xi) \in
    \{
    \texttt{raw\_memory},
    \texttt{processed\_memory},
    \texttt{raw\_skill},
    \texttt{delegated\_computation},
    \texttt{blocked}
    \}.
\end{equation}
These modes correspond to three abstract actions:
\begin{equation}
    a \in \{\mathrm{raw},\mathrm{processed},\mathrm{block}\}.
\end{equation}
The mapping depends on the asset type:
\[
\begin{array}{lll}
\text{memory} & \mathrm{raw} & \text{release raw or lightly redacted memory},\\
\text{memory} & \mathrm{processed} & \text{release summarized or sanitized memory},\\
\text{skill} & \mathrm{raw} & \text{release skill code},\\
\text{skill} & \mathrm{processed} & \text{execute locally and return only output}.
\end{array}
\]
The key asymmetry is that processed memory still exports transformed state,
whereas processed skill exchange exports behavior without exposing the skill
implementation.

\noindent\textbf{Outbound safety evaluation.}
The outbound safety backend evaluates each candidate using asset sensitivity,
skill risk, requester trust, authorization policy, and transformation
requirements. A high-trust requester may receive raw low-risk assets, while
sensitive assets may require redaction, summarization, or forced delegation.
Assets that violate the outbound policy are blocked:
\begin{equation}
    \mathrm{Allow}(\xi,A_i)=0.
\end{equation}
Before final transmission, generated or transformed payloads may pass through an
output guard to reduce leakage risk.

\noindent\textbf{Processed memory materialization.}
When the selected tier is \texttt{processed\_memory}, the responder materializes
a sanitized memory artifact. This may involve redaction, summarization, and a
final output-guard check. The resulting payload is intended to preserve
task-relevant information while removing unnecessary sensitive detail.

\noindent\textbf{Delegated computation materialization.}
When the selected tier is \texttt{delegated\_computation}, the responder keeps
the skill local. The exchange engine binds task arguments to the skill interface,
checks whether missing arguments can be clarified, executes the skill locally
when possible, and returns only the result. If the execution result is sensitive,
the result may be wrapped into a semantic answer rather than exposing raw output.

\noindent\textbf{Selection of exchange items.}
After relevance, trust, and safety filtering, the responder returns a bounded
set of exchange items:
\begin{equation}
    \mathcal{X}_{j\to i}
    =
    \operatorname{TopK}_{\xi\in\Xi_j}
    \mathrm{Score}(\xi),
\end{equation}
where the score reflects relevance, release tier, risk, and policy constraints.
The returned exchange decision records the selected items, release modes, and
any blocked or no-asset status.

\noindent\textbf{Inbound validation and exchange memory.}
The requester does not directly write received payloads into core work memory.
Each inbound payload is first validated under the receive-side safety policy.
Payloads suspected of injection, misinformation, malformed content, or policy
violation are rejected and may trigger a safety incident. Accepted payloads are
normalized into asset records and stored in exchange memory:
\begin{equation}
    E_i(t+1)=\mathrm{Integrate}(E_i(t),\mathcal{X}_{j\to i}),
\end{equation}
separate from the agent's work memory $W_i(t)$. Later retrieval may use both work
memory and exchange memory, but promotion into longer-lived memory remains a
separate decision.

\noindent\textbf{Safety incidents.}
If the receive-side gate detects a safety incident, the requester may write a
trust-update block to the evidence log and penalize the source peer. This links
exchange-time safety failures back into the evidence-derived trust mechanism used
by routing and future exchange decisions.

\noindent\textbf{Post-exchange feedback.}
After using the received payload, the requester can send feedback containing
answer-level quality, asset-level quality, per-asset comments, and a reason code
such as successful resolution, low quality, or capability mismatch. Depending on
the configured feedback policy, feedback may be sent for all exchanges, only
negative outcomes, or unresolved demands. The receiver uses this feedback for
local asset refinement, quality-score updates, and evidence persistence.

\noindent\textbf{Role in the protocol.}
\textsc{AEP-Exchange} realizes the admissible-context constraint of OPAC at
runtime. It does not simply transmit whatever a remote peer requests; it selects
task-relevant assets, applies trust and topology gates, chooses an appropriate
release mode, validates inbound content, separates received context from core
memory, and emits feedback for future coordination and routing.

\subsection{\textsc{AEP-Ledger} Operational Details}
\label{app:aep-ledger}

This appendix specifies the operational realization of \textsc{AEP-Ledger}. The
ledger is a verifiable evidence infrastructure: it stores signed protocol
evidence, verifies remote evidence blocks, derives queryable views for upper
subprotocols, and compresses old evidence through checkpoints. It does not
itself decide whether an asset should be released, which peer should be selected,
or how a demand should be negotiated.

\subsubsection{Ledger Details}

\noindent\textbf{Ledger realization.}
AEP supports two compatible ledger realizations. The first is a signed
hash-linked DAG, where each block hash commits to the block's canonical content
and signature. The second is a Merkle DAG, where the block hash is the Merkle
root over normalized header and payload fields. The Merkle variant supports
field-level proofs: a peer can verify that a specific field, such as a public key
or capability description, belongs to a committed block without receiving the
entire payload.

\noindent\textbf{Evidence block.}
The atomic evidence object is a block
\begin{equation}
    b =
    (h_b, \mathrm{parents}_b, a_b, \ell_b, s_b, p_b, \sigma_b),
\end{equation}
where $h_b$ is the block hash or Merkle root, $\mathrm{parents}_b$ is the set of
parent hashes, $a_b$ is the author identifier, $\ell_b$ is the logical clock,
$s_b$ is the per-author sequence number, $p_b$ is the payload, and $\sigma_b$ is
the author signature. The signed canonical content includes the sorted parent
hashes, author identity, logical clock, sequence number, and canonical payload:
\begin{equation}
    \mathrm{Canon}(b)
    =
    \mathrm{Canon}
    (\mathrm{parents}_b,a_b,\ell_b,s_b,p_b).
\end{equation}
The signature is therefore bound not only to the payload, but also to the local
DAG context in which the evidence is appended.

\noindent\textbf{Evidence payloads.}
A payload is typed by its \texttt{kind}. The implementation uses payload kinds
such as
\[
\begin{aligned}
&\texttt{capability\_announce},\quad
\texttt{tx\_claim},\quad
\texttt{rx\_claim},\\
&\texttt{topology\_probe\_tx},\quad
\texttt{topology\_probe\_rx},\quad
\texttt{trust\_update},\\
&\texttt{quality\_feedback},\quad
\texttt{checkpoint\_commit}.
\end{aligned}
\]
Capability announcements publish signed public metadata. Exchange and
topology-probe claims provide interaction evidence. Trust updates and quality
feedback provide later evidence for safety and service-quality derivation.
Checkpoint commits summarize finalized ledger state.

\noindent\textbf{Local append.}
When agent $A_i$ appends a local block, it chooses the current DAG heads as
parents, increments its per-author sequence number, signs the canonical block
content, and computes the block commitment. In the hash-linked variant, the block
hash is computed from the canonical content and signature. In the Merkle variant,
header and payload fields are committed as Merkle leaves and reduced to a root.
The appended block becomes a new DAG head, while previous heads referenced as
parents cease to be frontier heads.

\noindent\textbf{Remote ingestion.}
When a remote block is received, the ledger performs layered validation:
\begin{equation}
\mathrm{Accept}(b)=
\mathbf{1}\{
\mathrm{FreshHash}\wedge
\mathrm{NotBlacklisted}\wedge
\mathrm{HashValid}\wedge
\mathrm{SigValid}\wedge
\mathrm{SeqValid}\wedge
\mathrm{ParentsAvailable}
\}.
\end{equation}
Concretely, the receiver rejects duplicate hashes, rejects blocks authored by
blacklisted agents, recomputes the hash or Merkle root, verifies the author
signature, checks per-author sequence monotonicity, and confirms that parent
hashes are locally available or already covered by a finalized checkpoint. Blocks
with missing parents may be held as orphans until their ancestors arrive.

\noindent\textbf{Equivocation and blacklisting.}
The per-author sequence number gives a lightweight equivocation test. If two
distinct blocks share the same $(\mathrm{author},\mathrm{seq})$ but have
different block hashes, they constitute double-signing evidence:
\begin{equation}
    b \neq b',
    \quad
    a_b=a_{b'},
    \quad
    s_b=s_{b'}.
\end{equation}
When such evidence is detected, the author is blacklisted and its future blocks
are rejected. This makes blacklisting evidence-driven rather than reputation-only.

\noindent\textbf{Merkle field proofs.}
In the Merkle DAG realization, each normalized field is hashed into a leaf. A
field proof for key $k$ and value $v$ consists of the leaf hash and sibling hashes
on the path to the block root. Verification recomputes
\begin{equation}
    r = \mathrm{MerkleRoot}(k,v,\mathrm{proof})
\end{equation}
and accepts the field claim only if $r=h_b$. This allows a peer to prove selected
fields of a committed capability or checkpoint block without revealing the full
block body.

\noindent\textbf{Evidence-backed topology.}
The evidence-backed interaction topology $G_i^{\mathrm{ev}}(t)$ is inferred
from matched signed claims, not from routing-table hop counts. Let
$\mathrm{Tx}(a,b,h)$ denote a transmit claim by $a$ toward $b$ with interaction
hash $h$, and let $\mathrm{Rx}(b,a,h)$ denote the corresponding receive claim.
Then the ledger-derived topology contains an undirected edge only if both sides
are present:
\begin{equation}
    (A_a,A_b)\in G_i^{\mathrm{ev}}(t)
    \quad\Longleftrightarrow\quad
    \exists h:
    \mathrm{Tx}(a,b,h)\wedge \mathrm{Rx}(b,a,h).
\end{equation}
The same rule applies to matched topology-probe claims. This prevents a
gossip-propagated route from being treated as verified connectivity unless it is
backed by two-sided signed evidence.

\noindent\textbf{Queryable evidence view.}
Let $\mathcal{L}_i(t)$ denote the local evidence log of agent $A_i$. The ledger
exposes a derived query view
\begin{equation}
    \mathcal{Q}_i(t)=\mathrm{Query}(\mathcal{L}_i(t)),
\end{equation}
which contains verified capability records, blacklist status, matched interaction
claims, topology-probe evidence, trust-update evidence, quality-feedback records,
and checkpoint state. Upper subprotocols query $\mathcal{Q}_i(t)$ but do not
mutate it directly.
It combines signed
ledger facts with locally derived scores. A typical derived entry contains the
latest signed capability announcement for an author, its source block hash,
blacklist status, verified connectivity in $G_i^{\mathrm{ev}}(t)$, derived trust
score, service-quality estimate, and freshness metadata. Importantly, this view
does not trust raw gossip entries directly; it reconstructs peer state from
signed evidence plus local derivations.

\noindent\textbf{Trust and quality boundary.}
The ledger stores trust-update and quality-feedback evidence, but the final
ratings are derived outside the ledger. Routing may compute safety trust by
propagating local and distributed safety assessments over $G_i^{\mathrm{ev}}(t)$,
while service quality may be maintained as an EMA over quality-feedback evidence.
This preserves a clean boundary: \textsc{AEP-Ledger} stores and verifies evidence,
whereas upper subprotocols derive policy state from that evidence.

\noindent\textbf{Checkpoint construction.}
Long-running deployments accumulate many evidence blocks, so AEP periodically
constructs a checkpoint
\begin{equation}
    C_e =
    (
    e,\rho_e,\mathcal{H}_e,\mathcal{S}_e,\Sigma_e
    ),
\end{equation}
where $e$ is the epoch, $\rho_e$ is the checkpoint state root, $\mathcal{H}_e$
contains covered block hashes and covered DAG heads, $\mathcal{S}_e$ contains
state snapshots and proof references, and $\Sigma_e$ is quorum-backed signature
evidence. The committed state may include trust snapshots, quality snapshots,
the evidence-backed topology, validator state, capability references, topology
proof references, trust evidence references, and quality evidence references.
The state root is computed by hashing the canonical checkpoint content.

\noindent\textbf{Checkpoint modes and finalization.}
A checkpoint may be a genesis checkpoint when no previous finalized checkpoint
exists, a normal checkpoint extending a single compatible predecessor, or a merge
checkpoint combining multiple compatible checkpoint heads. We use cryptographic consensus mechanism to make sure the checkpoint compression process safe and verifiable~\citep{gifford1979weighted, yin2018hotstuff}. A designated proposer
builds the checkpoint candidate, validators verify its canonical content and
membership conditions, and the checkpoint is finalized once the accumulated
signature weight reaches the quorum threshold:
\begin{equation}
    \sum_{v\in \mathrm{Signers}(C_e)} w_v
    \geq
    \theta_{\mathrm{quorum}}.
\end{equation}
 The finalized checkpoint is then recorded as checkpoint evidence.

\noindent\textbf{Checkpoint application.}
When an agent applies a finalized checkpoint, it can restore derived state from
the committed snapshots. In particular, it may install the trust snapshot,
verified-topology snapshot, quality snapshot, and capability references, then
reconstruct local routing entries from checkpointed capability facts. This gives
new or recovering agents a fast path to a coherent evidence-derived state without
replaying the entire historical DAG.

\noindent\textbf{Checkpoint barriers and delta synchronization.}
After finalization, the ledger registers a checkpoint barrier consisting of the
covered block hashes, covered heads, and per-author maximum covered sequence
numbers. During later synchronization, export procedures can stop traversing
history once they reach a barrier and need only send post-checkpoint delta blocks.
A newly joined peer can therefore verify the latest finalized checkpoint root and
then download the evidence blocks that occur after it.

\noindent\textbf{Pruning.}
Checkpointing permits physical pruning without abandoning append-only evidence
semantics. Blocks covered by a finalized checkpoint can be removed from active
storage, while the checkpoint commit, recent post-checkpoint blocks, unresolved
claims, latest capability announcements, and proof-critical ancestor fragments
are retained as needed. After pruning, DAG heads and per-author sequence state
are recomputed subject to the checkpoint barrier.

\noindent\textbf{Persistence and recovery.}
On recovery from persistent storage, an agent reloads persisted heads, walks
backward through their ancestors, reinserts blocks in topological order, clears
orphan state where possible, and recomputes derived state such as blacklist
status, verified topology, trust, and quality. Since checkpoint barriers are
derived from finalized checkpoint evidence, recovery can rebuild them rather
than treating them as separate trusted state.

\noindent\textbf{Summary.}
\textsc{AEP-Ledger} provides durable evidence, not direct policy decisions. Its
append-only authenticated DAG records protocol claims; matched claims induce
verified topology; queryable views expose evidence-derived state to upper layers;
and quorum-backed checkpoints compress history while preserving verifiability.

\subsubsection{Clock Model and Runtime Time Semantics}

AEP uses three complementary notions of time. The first is a local step counter,
which is a monotone runtime counter advanced by the local scheduler. It is used
only to trigger periodic background tasks, including gossip rounds, trust
recomputation, knowledge maintenance, memory-summary refresh, skill-summary
refresh, and keyword-index refresh. Since the step counter is local to a single
agent runtime, it does not participate in distributed ordering and should not be
interpreted as a global tick or a physical-time measurement.

The second notion of time is the protocol logical clock. Each agent maintains a
Lamport-style logical clock. Local protocol events advance the clock, and inbound
messages carrying a logical-clock field update the receiver according to the
standard rule
\begin{equation}
    \ell_i \leftarrow \max\{\ell_i, \ell_{\mathrm{recv}}\} + 1 .
\end{equation}
This logical clock is the primary time base for evidence-bearing protocol state.
In particular, locally appended DAG blocks record the current logical clock,
flow entries store the logical time at which they were last observed, gossip
export filters and stale-entry decay are evaluated against logical freshness,
sessions are pruned by logical timeout intervals, checkpoints are proposed at
logical-clock intervals, and emitted protocol events carry logical-clock
metadata. As a result, any evidence generated by an agent after receiving remote
protocol messages is logically ordered after the observed remote events.

The third notion of time is physical millisecond time. AEP uses physical time
only for network-facing asynchronous timeout semantics, such as detecting whether
a semantic exchange request has exceeded its wall-clock deadline. This is
necessary because asynchronous exchange interacts with real transport behavior,
including RPC latency, socket timeouts, retries, and partitions. Physical time is
therefore an implementation-level timeout signal rather than a source of
distributed causality.

During normal execution, the local step counter determines when a periodic task
should be attempted. When such a task generates protocol evidence or sends a
protocol message, the logical clock is advanced and the resulting evidence is
stamped with logical time. During recovery, the agent restores the logical clock
to at least the maximum logical clock observed in the local ledger, aligns the
local step counter to avoid immediate background-task bursts, and resets interval
trackers to the restored clock values. This prevents recovered agents from
appending evidence that appears causally earlier than already persisted ledger
state.

\begin{wraptable}{r}{0.5\textwidth}
\vspace{-1.5em}
\centering
\captionsetup{font=small}
\caption{Safety Comparison between AEP and Shared Memory Collaboration}
\label{tab:aep_shared_memory_safety}
\resizebox{\linewidth}{!}{%
\begin{tabular}{lcc}
\toprule
Protocol & Leakage Rate & Injection Success Rate \\
\midrule
Shared Memory & 100.00\% & 100.00\% \\
AEP           & 0.00\%   & 0.00\%   \\
\bottomrule
\end{tabular}
}
\end{wraptable}

\section{Extended Experiment and Extra Experiment Details}

\subsection{Safety Comparison to Shared Memory Paradigms}
We compare shared memory paradigms and our protocol's robustness to injection attack and privilege extraction attack. As shown in Table.~\ref{tab:aep_shared_memory_safety}, under the full shared memory paradigm cooperation incurs a 100\% privacy leakage rate and injection success rate, while our protocols defense all the attacks.

\subsection{Robustness to Scalable Attacks under Multi-Agent Scenario}
\label{app:robustness_scalable_attacks}

\noindent\textbf{Experimental setting.}
We evaluate whether AEP remains robust when the fraction of malicious agents increases in a multi-agent collaboration network.
The experiment is conducted on the Coding domain under a middle-scale network, where task-relevant information and capabilities are distributed across agents.
For each protocol, we vary the malicious-agent ratio as
\(
\rho \in \{0, 0.1, 0.4\},
\)
where $\rho$ denotes the fraction of agents controlled by the attacker.
The clean setting $\rho=0$ is used as the reference condition for computing task score degradation, while the attacked settings $\rho=0.1$ and $\rho=0.4$ are reported in Table~\ref{tab:scalable_attack_robustness}. As a controlled defense baseline, we further augment A2A and Broadcast with the same external Qwen3Guard-Gen-0.6B filter for inbound and outbound messages, while AEP relies solely on its native safety and exchange controls.

\noindent\textbf{Attack design.}
Inspired by prior work~\cite{amayuelas2024multiagent,cui2025mad, dong2025memory,chen2024agentpoison}, we instantiate a scalable poisoning attack where malicious agents behave as ordinary collaborators during discovery, but return poisoned payloads once contacted.
Unlike the explicit injection attacks considered earlier, the payloads contain no overtly malicious instructions or policy-violating content.
Instead, they are \emph{benign-looking but factually incorrect or misleading} claims, faked intermediate results, or subtly corrupted evidence, which cannot be reliably removed by direct safety gating.

\noindent\textbf{Metrics.}
We report two primary metrics.
First, normalized task score drop, denoted as $\mathrm{nTSD}$ (normalized Task Score Drop), measures end-to-end utility degradation:
\begin{equation}
    \mathrm{nTSD}
=
\frac{\mathrm{origin~task~score} - \mathrm{attacked~task~score}}
{\max(\mathrm{origin~task~score}, \epsilon)}.
\end{equation}
A lower $\mathrm{nTSD}$ means that the protocol preserves task quality better under attack.
Second, we report the change in $\mathrm{MCR}$ (Malicious Contact Rate) from early to late epochs, written as
\[
\mathrm{MCR}_{\mathrm{early}} \rightarrow \mathrm{MCR}_{\mathrm{late}}.
\]
Here, $\mathrm{MCR}$ measures the fraction of peer contacts that involve malicious agents.
This metric evaluates whether a protocol can reduce malicious exposure over time.

\subsection{External-Guard Baselines and Scalable Attack}
\label{app:guarded-added}

\noindent\textbf{Controlled defense configuration and calibration.}
We augment A2A and Broadcast with the same external defense, denoted A2A + Guard and Broadcast + Guard, using Qwen3Guard-Gen-0.6B for inbound and outbound screening at the corresponding communication boundaries. The guard follows its official chat template without a customized safety policy, while the backbone, tasks, topology, and attack instances are otherwise held fixed. AEP instead uses its native release-side and receive-side controls and is not equipped with the external Guard.
We calibrate the guard on an open-source prompt-injection dataset\footnote{Available at \url{https://huggingface.co/datasets/neuralchemy/Prompt-injection-dataset}.}. A deterministic calibration split is drawn from the validation set and a disjoint evaluation split from the test set, each containing 200 benign and 200 attack samples. For each message, we define the risk score as $r=1-P(\texttt{Safe})$, where probabilities are normalized over the constrained \texttt{Safe}, \texttt{Unsafe}, and \texttt{Controversial} logits. Inbound and outbound thresholds are calibrated independently to maximize attack recall subject to an empirical benign false-positive rate of at most $1\%$ on the calibration split. 

\noindent\textbf{Safety results.}
Table~\ref{tab:guard-safety-added} evaluates direct and paraphrased injection attacks together with privacy-extraction probes. All methods have zero observed privacy leakage in this guarded comparison. The external Guard does not prevent the tested injected payloads from entering the downstream workflows of A2A and Broadcast, whose injection success rates remain 100\%. AEP records 0\% injection success in Coding and Daily-Service and $2.90$\% in Research. These results distinguish protection against unauthorized release from protection against accepting malicious context; neither metric alone establishes whether a final answer was behaviorally hijacked.

\noindent\textbf{Scalable multi-agent attack results.}
We also evaluate the guarded baselines on the middle-scale Coding under malicious agent setup of Section~\ref{sec:robustness_scalable_attacks}, where contacted malicious peers return benign-looking but incorrect claims or intermediate results. Table~\ref{tab:guard-robustness-added} reports normalized task-score drop (nTSD) relative to the corresponding clean condition and early-to-late malicious contact rate (MCR). At $\rho=0.4$, the Guard reduces A2A's nTSD from 1.000 to 0.500 and Broadcast's from 0.350 to 0.300, but their malicious-contact rates remain unchanged. At $\rho=0.1$, A2A + Guard instead has a larger drop than unguarded A2A (0.357 versus 0.017), so the effect on utility is not uniformly beneficial. AEP records nTSD values of 0.000 and 0.035 at the two attack ratios and reduces late-stage MCR to zero in both cases. Content screening does not itself update the baseline collaborator-selection policies; AEP additionally uses recorded interaction outcomes in subsequent routing and exchange decisions.

\begin{table*}[t]
\centering

\begin{minipage}[t]{0.48\textwidth}
\centering
\small

\captionof{table}{
Safety evaluation with guarded A2A and Broadcast under middle scale network.
}
\label{tab:guard-safety-added}

\vspace{0.3em}

\resizebox{\linewidth}{!}{%
\begin{tabular}{llrr}
\toprule
Scenario 
& Protocol 
& Injection Success $\downarrow$ 
& Privacy Leakage $\downarrow$ \\
\midrule

Coding 
& A2A + Guard 
& $100.00 \pm 0.00$ 
& $0.00 \pm 0.00$ \\
& Broadcast + Guard 
& $100.00 \pm 0.00$ 
& $0.00 \pm 0.00$ \\
& AEP 
& $0.00 \pm 0.00$ 
& $0.00 \pm 0.00$ \\

\midrule

Daily-Service 
& A2A + Guard 
& $100.00 \pm 0.00$ 
& $0.00 \pm 0.00$ \\
& Broadcast + Guard 
& $100.00 \pm 0.00$ 
& $0.00 \pm 0.00$ \\
& AEP 
& $0.00 \pm 0.00$ 
& $0.00 \pm 0.00$ \\

\midrule

Research 
& A2A + Guard 
& $100.00 \pm 0.00$ 
& $0.00 \pm 0.00$ \\
& Broadcast + Guard 
& $100.00 \pm 0.00$ 
& $0.00 \pm 0.00$ \\
& AEP 
& $2.90 \pm 1.91$ 
& $0.00 \pm 0.00$ \\

\bottomrule
\end{tabular}%
}

\end{minipage}
\hfill
\begin{minipage}[t]{0.48\textwidth}
\centering
\small

\captionof{table}{
Robustness under scalable malicious agent poisoning on middle-scale Coding.
}
\label{tab:guard-robustness-added}

\vspace{0.3em}

\resizebox{\linewidth}{!}{%
\begin{tabular}{lrrc}
\toprule
Protocol 
& Malicious Ratio $\rho$ 
& nTSD $\downarrow$ 
& $\mathrm{MCR}_{\mathrm{early}}
\rightarrow
\mathrm{MCR}_{\mathrm{late}}$ $\downarrow$ \\
\midrule

A2A 
& 0.1 
& 0.017 
& $0.333 \rightarrow 0.333$ \\
A2A 
& 0.4 
& 1.000 
& $0.333 \rightarrow 0.333$ \\

\midrule

A2A + Guard 
& 0.1 
& 0.357 
& $0.333 \rightarrow 0.333$ \\
A2A + Guard 
& 0.4 
& 0.500 
& $0.333 \rightarrow 0.333$ \\

\midrule

Broadcast 
& 0.1 
& 0.000 
& $0.133 \rightarrow 0.133$ \\
Broadcast 
& 0.4 
& 0.350 
& $0.400 \rightarrow 0.400$ \\

\midrule

Broadcast + Guard 
& 0.1 
& 0.000 
& $0.133 \rightarrow 0.133$ \\
Broadcast + Guard 
& 0.4 
& 0.300 
& $0.400 \rightarrow 0.400$ \\

\midrule

AEP 
& 0.1 
& 0.000 
& $0.167 \rightarrow 0.000$ \\
AEP 
& 0.4 
& 0.035 
& $0.500 \rightarrow 0.000$ \\

\bottomrule
\end{tabular}%
}

\end{minipage}

\end{table*}

\subsection{Cross-Model Evaluation}
\label{app:cross-model-added}

We replace the backbone while retaining the same middle-scale Coding tasks, agent initialization, physical topology, protocol configuration, and evaluation procedure. Table~\ref{tab:cross-model-added} reports the resulting utility and token cost. All three backbones complete every evaluated task. GPT-5.5 achieves a Task Score close to DeepSeek-V4-Flash at a similar token cost, while Qwen3.7-Max yields lower answer quality and moderately higher token consumption. 

\begin{table}[htbp]
\centering
\small
\caption{Cross-model evaluation of AEP on Coding in the middle-scale network. Only the backbone is replaced; tasks, agent initialization, topology, protocol configuration, and evaluation procedure are held fixed.}
\label{tab:cross-model-added}
\resizebox{0.65\linewidth}{!}{%
\begin{tabular}{lrrr}
\toprule
Backbone & Task Score $\uparrow$ & Completion (\%) $\uparrow$ & Tokens/task (k) $\downarrow$ \\
\midrule
DeepSeek-V4-Flash & 99.17 & 100.00 & 10.1 \\
GPT-5.5 & 99.67 & 100.00 & 10.4 \\
Qwen3.7-Max & 83.33 & 100.00 & 11.2 \\
\bottomrule
\end{tabular}
}

\end{table}

\subsection{Implementation Details of Production-Grade Protocols}
\label{app:sdk-added}

We execute the production-grade protocols with upstream SDK packages \texttt{anp} (0.9.3)\footnote{ANP: \url{https://www.agent-network-protocol.com/specs/1.2/white-paper}}, \texttt{agora-protocol} (0.2.0)\footnote{Agora: \url{https://agoraprotocol.org/}}, and \texttt{acp-sdk} (1.0.3)\footnote{ACP: \url{https://agentcommunicationprotocol.dev/introduction/welcome}}. 
The protocol request validation, serialization, discovery responses, and task handling execute within their applications, while operating-system HTTP transport is redirected through the benchmark's simulated network for common reachability, failure, latency, and byte accounting. 
To address Agora's limited agent-discovery capability, we additionally augment it with direct-neighbor gossip for exchanging routing cards, enabling a fairer comparison.

\subsection{Metric Definitions}
\label{app:metric-definitions}

We evaluate AEP and all baselines along three dimensions: task utility, safety, and robustness.
Let $\mathcal{D}$ denote the evaluation task set, and let $N=|\mathcal{D}|$.
For each task $x_t \in \mathcal{D}$, a protocol produces an output $\hat{y}_t$, an interaction trace $\mathcal{E}_t$, and a set of collaboration contacts $\mathcal{C}_t$.

\noindent\textbf{Task Score.}
Task Score measures the end-to-end quality of the final answer.
For each task, an LLM judge assigns a rubric-based score
\[
s_t = J(x_t, \hat{y}_t, y_t^\star) \in [0,100],
\]
where $y_t^\star$ is the reference answer or task-specific grading target.
The reported Task Score is the average score over all tasks:
\[
\mathrm{TaskScore}
=
\frac{1}{N}\sum_{t=1}^{N} s_t .
\]
Higher values indicate better task performance.

\noindent\textbf{Completion Rate.}
Completion Rate measures whether a protocol can finish the task within the interaction and time budget.
Let $c_t \in \{0,1\}$ indicate whether task $t$ returns a valid final answer before the budget is exhausted.
The completion rate is
\[
\mathrm{CompletionRate}
=
\frac{1}{N}\sum_{t=1}^{N} c_t .
\]
We report this value as a percentage.

\noindent\textbf{Token Cost.}
Token cost measures the average communication and reasoning cost per task.
For each task, we count all tokens consumed by LLM calls and collaborative communication events, including prompts, generated responses, negotiation messages, exchange requests, and returned payloads.
Let $\mathrm{tok}(e)$ be the token count of event $e \in \mathcal{E}_t$.
The average token cost is
\[
\mathrm{Tokens/task}
=
\frac{1}{N}\sum_{t=1}^{N}\sum_{e\in \mathcal{E}_t} \mathrm{tok}(e).
\]
In the main tables, we report this value in thousands of tokens.

\noindent\textbf{Privacy Leakage Rate.}
Privacy Leakage Rate measures how often protected private information is exposed to unauthorized parties or final outputs.
For each task, let $\mathcal{P}_t$ be the set of protected private items, and let $\ell(p)=1$ indicate that protected item $p$ is disclosed to an unauthorized agent, tool call, message, or final answer.
The leakage rate is
\[
\mathrm{LeakageRate}
=
\frac{\sum_{t=1}^{N}\sum_{p\in \mathcal{P}_t}\mathbb{I}[\ell(p)=1]}
{\sum_{t=1}^{N}|\mathcal{P}_t|}.
\]
We count each protected item at most once per task, even if it appears multiple times in the same interaction trace.
Lower values indicate stronger privacy protection.

\noindent\textbf{Injection Success Rate.}
Injection Success Rate measures the fraction of adversarial injection attempts that are accepted into the downstream workflow.
Let $\mathcal{A}_t$ be the set of malicious injection attempts in task $t$, and let $a$ be successful if the injected content is accepted into the agent's reasoning context, exchange memory, tool-use plan, delegated workflow, or final response.
The injection success rate is
\[
\mathrm{InjectionSuccessRate}
=
\frac{\sum_{t=1}^{N}\sum_{a\in \mathcal{A}_t}\mathbb{I}[\mathrm{success}(a)=1]}
{\sum_{t=1}^{N}|\mathcal{A}_t|}.
\]
Lower values indicate stronger robustness against malicious external context.

\noindent\textbf{Normalized Task Score Drop.}
For robustness experiments with malicious agents, we measure utility degradation relative to the clean setting.
Let $\mathrm{Score}_{0}$ be the Task Score under no attack, and let $\mathrm{Score}_{\rho}$ be the Task Score when the malicious-agent ratio is $\rho$.
The normalized task score drop is
\[
\mathrm{nTSD}(\rho)
=
\frac{\mathrm{Score}_{0}-\mathrm{Score}_{\rho}}
{\max(\mathrm{Score}_{0},\epsilon)},
\]
where $\epsilon>0$ is a small constant for numerical stability.
Lower values indicate that the protocol preserves task quality better under attack.

\noindent\textbf{Malicious Contact Rate.}
Malicious Contact Rate measures how frequently benign agents interact with malicious agents.
Let $\mathcal{C}$ be the set of collaboration contacts in a given evaluation window, and let $\mathrm{mal}(j)=1$ indicate that the contacted peer $j$ is malicious.
We define
\[
\mathrm{MCR}
=
\frac{\sum_{(i,j)\in \mathcal{C}}\mathbb{I}[\mathrm{mal}(j)=1]}
{|\mathcal{C}|}.
\]
For long-term robustness experiments, we report $\mathrm{MCR}_{\mathrm{early}} \rightarrow \mathrm{MCR}_{\mathrm{late}}$ to show whether the protocol reduces malicious exposure over repeated interactions.

\section{Extended OPAC Formulation}
\label{app:opac-full}

This appendix records the full version of Section~\ref{sec:OPAC}, while the main text keeps only the compact formulation needed to support the protocol design and theorems in Section~3.

We formalize OPAC as a distributed multi-objective decision problem over open private-agent networks. Let
\begin{equation}
\mathcal{V} = \{A_1, A_2, \dots, A_n\}
\end{equation}
denote a collection of private agents embedded in an unknown physical communication graph
\begin{equation}
G = (\mathcal{V}, \mathcal{E}).
\end{equation}
For theoretical analysis, we also use a weighted matrix representation $W^{\mathrm{phy}} \in \mathbb{R}^{n \times n}$ of the physical topology when convenient. Unlike centralized or fully observable multi-agent systems, no agent is assumed to know $G$ globally. Instead, each agent $A_i$ maintains a time-dependent local cognitive map
\begin{equation}
G_i(t) = (V_i(t), E_i(t)),
\end{equation}
together with a weighted cognitive topology $W_i^{\mathrm{cog}}(t)$ and a routing state $R_i(t)$ derived from currently available metadata.

Each agent $A_i$ is associated with a task variable pair $(X_i, Y_i)$, a private memory store $M_i$, and a local auditing policy governing what information may be received, retained, or transmitted. The goal of cooperation is to improve prediction or decision quality for $Y_i$ by augmenting local context with externally acquired context assets while respecting privacy, trust, and communication constraints.

\subsection{Context Assets, Collaboration Trajectories, and Policies}

The basic unit of cooperation in OPAC is not a full response, but a \emph{context asset}. We use this term abstractly to denote a transferable unit of task-relevant context that can be exchanged and integrated into another agent's local reasoning state. Context assets include memory fragments, sanitized summaries, skill artifacts, and delegated execution outputs; a memory slot is an important special case. Unlike a fully generated response, a context asset is intended to be reusable across tasks and amenable to asynchronous exchange. Section~\ref{sec:aep-exchange} later refines this operationally, while the theory continues to track all admitted contributions through the same imported-context abstraction.

For agent $A_i$, a cooperation event under policy $\pi$ is written as
\begin{equation}
e_k = (A_j \rightarrow A_i,\, a_k,\, m_{j \rightarrow i}^{(k)}),
\end{equation}
where $a_k \in \{\texttt{raw}, \texttt{processed}, \texttt{block}\}$ records the exchange action selected for that interaction and $m_{j \rightarrow i}^{(k)}$ denotes the admitted context asset for that event. A collaboration trajectory is then
\begin{equation}
\Gamma_i^{\pi}(t) = (e_1, e_2, \dots, e_K),
\end{equation}
and the aggregated external context delivered to $A_i$ is denoted by
\begin{equation}
m_{\rightarrow i}^{\pi}(t) = \{m_{j \rightarrow i}^{(k)}\}_{k=1}^{K}.
\end{equation}
The effective context available to $A_i$ under policy $\pi$ is
\begin{equation}
\widetilde{M}_i^{\pi}(t) = M_i \cup m_{\rightarrow i}^{\pi}(t).
\end{equation}

A local cooperation policy $\pi_i$ maps the local state of agent $A_i$---including its observation, private memory, cognitive topology, routing state, and local safety policy---to routing, exchange, auditing, and integration actions. A network-level cooperation policy is denoted by
\begin{equation}
\pi = (\pi_1, \dots, \pi_n).
\end{equation}

\subsection{Safe Feasible Policy Class Under Partial Observability}

Because agents only have partial topology visibility, a policy can only act on candidates that are currently discoverable under the agent's local cognitive state. Let
\begin{equation}
\mathcal{A}_i(t)
\end{equation}
denote the admissible set of external context assets for agent $A_i$ at time $t$, where admissibility jointly requires discoverability, reachability, authorization, and successful auditing. Operationally, $\mathcal{A}_i(t)$ should be read as the agent's currently certified admissible set under local routing and audit state rather than as an oracle revelation of all truly safe transfers.

Let $\mathcal{S}$ denote the set of sensitive information types, and let
\begin{equation}
\mathrm{Auth}(s) \subseteq \mathcal{V}
\end{equation}
denote the agents authorized to access sensitive information $s \in \mathcal{S}$. For a joint policy $\pi$, we define the privacy leakage risk of agent $A_i$ as
\begin{equation}
R_i^{\mathrm{priv}}(\pi)
=
\max_{s \in \mathcal{S},\, A_i \notin \mathrm{Auth}(s)}
I\!\left(m_{\rightarrow i}^{\pi}(t); s\right).
\end{equation}
More generally, harmful-peer isolation, poisoning resistance, and content filtering are all absorbed into the admissible set and the local auditing policy.

Because the imported context $m_{\rightarrow i}^{\pi}(t)$ and the privacy risk $R_i^{\mathrm{priv}}(\pi)$ inherently depend on the release actions of other agents, the network-level safe feasible set cannot be decomposed into an independent Cartesian product. While the base action space $\Pi_1 \times \cdots \times \Pi_n$ is decomposable, the network-level joint safe policy set at time $t$ is defined collectively as:
\begin{equation}
\Pi_{\mathrm{safe}}(t) = \left\{ \pi \in \Pi_1 \times \cdots \times \Pi_n \;\middle|\; \forall i, \; m_{\rightarrow i}^{\pi}(t) \subseteq \mathcal{A}_i(t), \; R_i^{\mathrm{priv}}(\pi) \le \epsilon_i \right\}
\end{equation}
for task- or agent-dependent tolerances $\epsilon_i \ge 0$. This joint feasible class is the decision space of OPAC.

\subsection{Multi-Objective Evaluation of Cooperation Policies}

For theoretical analysis, we evaluate predictive performance under the Bayes-optimal log score. For a safe policy $\pi \in \Pi_{\mathrm{safe}}(t)$, define the local predictive utility of agent $A_i$ as
\begin{equation}
U_i(\pi)
=
\mathbb{E}\!\left[\log p^{\star}\!\left(Y_i \mid X_i, \widetilde{M}_i^{\pi}(t)\right)\right],
\end{equation}
where $p^{\star}$ denotes the Bayes-optimal predictive distribution. The aggregate network utility is then
\begin{equation}
U(\pi) = \sum_{i=1}^{n} U_i(\pi).
\end{equation}
This Bayes-optimal log score is used as an analytical surrogate for task utility; in empirical agent tasks, $Y_i$ may correspond to a reference answer, judge rubric, executable success criterion, or user-specified task outcome. Because utility is defined through predictive log score, Theorem~\ref{thm:utility-info} connects it to the task-relevant information carried by imported context assets.

In addition to utility, OPAC tracks a safety-trust objective and a resource objective. Let
\begin{equation}
T_i(\pi) = \mathbb{E}\!\left[\tau_i\!\left(\Gamma_i^{\pi}(t), m_{\rightarrow i}^{\pi}(t)\right)\right]
\end{equation}
denote a cooperation trust score, where $\tau_i(\cdot)$ rewards authorization compliance, routing through trusted peers, and avoidance of harmful or poorly sanitized transfers. Let
\begin{equation}
C_i(\pi) = C_i\!\left(\Gamma_i^{\pi}(t)\right)
\end{equation}
denote the communication and computation cost of the policy, including route discovery, transmission, sanitization, summarization, and local reasoning overhead. We aggregate these objectives as
\begin{equation}
U(\pi) = \sum_{i=1}^{n} U_i(\pi),
\qquad
T(\pi) = \sum_{i=1}^{n} T_i(\pi),
\qquad
C(\pi) = \sum_{i=1}^{n} C_i(\pi).
\end{equation}
Thus the hard-safe feasible class specifies which policies are allowable at all, while $T(\pi)$ distinguishes the more or less trustworthy operating points within that allowable set.

For comparison with the full-information ideal, let $\widetilde{M}_i^{\star}$ denote the effective context that agent $A_i$ would access under full global awareness and unconstrained discovery. The corresponding information-theoretic upper bound is
\begin{equation}
I_i^{\mathrm{global}} = I\!\left(Y_i ; \widetilde{M}_i^{\star} \mid X_i, M_i\right),
\end{equation}
which will later be useful for discussing topology-aware approximation gaps.

\subsection{Pareto Formulation of OPAC}

We represent each safe cooperation policy by the objective vector
\begin{equation}
z(\pi) = \big(U(\pi),\, T(\pi),\, -C(\pi)\big),
\end{equation}
so that larger coordinates are always preferred. For two safe policies $\pi, \pi' \in \Pi_{\mathrm{safe}}(t)$, we say that $\pi'$ \emph{dominates} $\pi$ if $z(\pi') \ge z(\pi)$ in coordinate wise and at least one inequality is strict.

The OPAC Pareto frontier at time $t$ is therefore
\begin{equation}
\mathcal{P}^{\star}(t)
=
\left\{
\pi \in \Pi_{\mathrm{safe}}(t)
\;\middle|\;
\nexists\, \pi' \in \Pi_{\mathrm{safe}}(t)
\text{ such that } z(\pi') \succ z(\pi)
\right\}.
\end{equation}
Accordingly, the goal of OPAC is not to identify a single policy under a fixed scalarization, but to characterize or approximate the safe Pareto frontier using only local observations, partial topology beliefs, and audited cooperation. This avoids baking in a universal operating point even though different deployments may prioritize privacy, latency, or utility differently.

In practice, agents only observe empirical estimates of these objectives from finitely many interaction samples. Let $\widehat{z}_t(\pi)$ denote the empirical estimate of $z(\pi)$ after $t$ observations. A solver for OPAC must therefore address two coupled challenges: it must expand the safe feasible set available under partial observability, and it must drive the empirical frontier induced by feasible policies toward the population Pareto frontier $\mathcal{P}^{\star}(t)$.

\section{Full Theoretical Results and Proofs}
\label{app:aep-theory-results}

This appendix records the idealized assumptions and full supporting formal statements behind the main theorem in Section~\ref{sec:aep-solver}. The main text keeps only the solver-level statement for page budget and exposition.

\subsection{Analysis Assumptions}
\label{app:aep-analysis-assumptions}

To avoid hidden assumptions, we collect the analytical conditions used by the appendix results in one place. For the consistency results, fix an evaluation snapshot at horizon $t$.
The task distribution, authorization rules, and reference physical
topology are held fixed during evaluation. Policy classes and population
objective maps may depend on the preceding history, but are fixed before
collecting fresh evaluation samples.
Within these consistency results, all objectives are maximized, and
$>$ denotes strict inequality in every coordinate. For a nonempty
finite set $Q\subset\mathbb{R}^3$, define
\[
\operatorname{PF}(Q)
=
\{x\in Q:\nexists y\in Q,\ y\ge x,\ y\ne x\},
\]
and
\[
g_Q(x)
=
\max_{y\in Q}\min_{1\le k\le3}(y_k-x_k).
\]
Throughout this appendix, $d_H$ is the Hausdorff distance induced by
$\|\cdot\|_\infty$.
\begin{itemize}
    \item[(A1)] \textbf{Bounded pruning losses.} For each routed candidate queue, the pruning loss variables $\Delta U_k$ are bounded in $[0,L]$ and are independent or otherwise satisfy the same concentration inequality used below.
    \item[(A2)] \textbf{Connected normalized physical topology.} The physical interaction graph is connected with spectral gap $\gamma^{\mathrm{phy}} > 0$, and semantic heterogeneity enters through a bounded bias term $b(\Gamma)$ together with bounded observation noise $\sigma$.
    \item[(A3)] \textbf{Topology-learning contraction.} The cognitive topology update contracts toward the physical graph in expectation according to
    \begin{equation}
    \mathbb{E}\bigl\|W_i^{\mathrm{cog}}(t+1)-W^{\mathrm{phy}}\bigr\|_2
    \le
    (1-\eta\gamma^{\mathrm{phy}})\,\mathbb{E}\bigl\|W_i^{\mathrm{cog}}(t)-W^{\mathrm{phy}}\bigr\|_2
    + \eta(b(\Gamma)+\sigma),
    \end{equation}
    where $\eta > 0$ is a learning step size satisfying $1-\eta\gamma^{\mathrm{phy}} \in (0,1)$.
    \item[(A4)] \textbf{Monotone admissible discovery.} In the idealized analysis model, improving routing evidence monotonically expands the relevant admissible discoverable set, so topology-constrained imported information is nondecreasing in expectation.
    \item[(A5)] \textbf{Certified sanitization and blocking.} The sanitization map $S$ is assumed to provide a certified residual leakage bound $\rho_S$, such that $I(S(\xi); s) \le \rho_S$. Transmitted assets retain at most this residual information, and perfect blocking implies $\rho_S = 0$.
    \item[(A6)] \textbf{Finite-entropy Bayes utility.} For each agent, the target variable $Y_i$ has finite conditional entropy given $(X_i,M_i)$, and predictive utility is measured by the Bayes-optimal log score from Section~\ref{sec:OPAC}.
    \item[(A7)] \textbf{Finite joint safe policy class.}
At horizon $t$, the reference class $\Pi_{\mathrm{safe}}(t)$ is a
nonempty finite set satisfying the jointly coupled safety constraints.

\item[(A8)] \textbf{Fresh bounded objective samples.}
For each policy in a class being evaluated, each objective estimate
is an average of $N_t$ fresh samples. Conditional on the evaluation
snapshot, the samples for each policy and coordinate are independent,
have the corresponding population objective as their common mean,
and lie in an interval of length at most $B_{\max}$.
This applies to both the reference map $z$ and the realized map $z_t$
introduced in A11. Independence across policies or coordinates is
not required. The count $N_t$ is per policy, not the total number
of interactions across the network. This bounded-sample condition
is additional to the finite-entropy condition in A6.

\item[(A9)] \textbf{Uniform topology sensitivity.}
Write $\mathbf W=(W_1,\ldots,W_n)$ and
\[
\|\mathbf W-\mathbf W'\|_{\max,2}
=
\max_i\|W_i-W_i'\|_2.
\]
For all relevant topology states and all
$\pi\in\Pi_{\mathrm{safe}}(t)$,
\[
\|z(\pi;\mathbf W)-z(\pi;\mathbf W')\|_\infty
\le
L_{\mathrm{obj}}\|\mathbf W-\mathbf W'\|_{\max,2}.
\]
Here $\|\cdot\|_2$ is the spectral norm.
Let
$\mathbf W^{\mathrm{phy}}
=(W^{\mathrm{phy}},\ldots,W^{\mathrm{phy}})$
and use $z(\pi)=z(\pi;\mathbf W^{\mathrm{phy}})$ for the reference
objective in the consistency comparison.

\item[(A10)] \textbf{Uniform dominance-gap error bound.}
Define
\[
Z_0(t)=\{z(\pi):\pi\in\Pi_{\mathrm{safe}}(t)\},
\qquad
Z_t=\{z_t(a):a\in\mathcal K_t\},
\]
where $\mathcal K_t$ and $z_t$ are specified in A11.
There is a constant $\kappa>0$, independent of $t$ and bounded
away from zero along any asymptotic regime considered, such that
\[
g_Q(x)
\ge
\kappa\,\operatorname{dist}_\infty
\bigl(x,\operatorname{PF}(Q)\bigr),
\qquad
x\in Q,\quad Q\in\{Z_0(t),Z_t\}.
\]
This condition is imposed on both population objective sets,
not on their noisy empirical estimates.

\item[(A11)] \textbf{Uniform safe realizability and empirical archive.}
At each snapshot, AEP induces a nonempty finite class $\mathcal K_t$
of safe realized policies with population objective map $z_t$.
There is a surjection
$R_t:\Pi_{\mathrm{safe}}(t)\to\mathcal K_t$ such that
\[
\sup_{\pi\in\Pi_{\mathrm{safe}}(t)}
\left\|
z_t(R_t(\pi))
-z(\pi;\mathbf W^{\mathrm{cog}}(t))
\right\|_\infty
\le
C_{\mathrm a}\,\overline{\Delta U},
\]
where
$\mathbf W^{\mathrm{cog}}(t)
=(W_1^{\mathrm{cog}}(t),\ldots,W_n^{\mathrm{cog}}(t))$,
$C_{\mathrm a}$ is independent of $t$, and
\[
\overline{\Delta U}
=
K_{\max}^{-1}\sum_{k=1}^{K_{\max}}\Delta U_k
\]
is the nonnegative loss in Lemma~\ref{thm:pruning}.
Every policy in $\mathcal K_t$ is evaluated as in A8.
The analyzed AEP archive retains exactly those policies that are
not exceeded in every empirical objective by more than
$2\varepsilon_t$ by another policy in $\mathcal K_t$.

\item[(A12)] \textbf{Evaluation after the initial topology transient.}
A3 holds for all $n$ agents on the fixed physical topology, with
$q=1-\eta\gamma^{\mathrm{phy}}\in(0,1)$ and
\[
e_0
=
\max_i\mathbb E
\|W_i^{\mathrm{cog}}(0)-W^{\mathrm{phy}}\|_2
<\infty.
\]
For Theorem~\ref{thm:aep-main}, let $M_t$ be a deterministic upper
bound on $|\Pi_{\mathrm{safe}}(t)|$ and define, for $0<\delta<1$,
\[
\varepsilon_t
=
B_{\max}
\sqrt{\frac{\log(12M_t/\delta)}{2N_t}}.
\]
The stated bound applies at evaluation horizons satisfying
\[
L_{\mathrm{obj}}q^t e_0\le\varepsilon_t.
\]
\end{itemize}

\subsection{Discovery Results}

\begin{theorem}[Stationary cognitive-topology error]
\label{thm:spectral-topology}
Under Assumption~A3, with step size $\eta \in (0, 1/\gamma^{\mathrm{phy}})$,
\begin{equation}
\limsup_{t\to\infty}\mathbb{E}\!\left\|W^{\mathrm{cog}}_i(t) - W^{\mathrm{phy}}\right\|_2
\le \frac{b(\Gamma)+\sigma}{\gamma^{\mathrm{phy}}}.
\end{equation}
\end{theorem}

\begin{lemma}[Concentration of auditing-induced pruning loss]
\label{thm:pruning}
Under the bounded independent pruning-loss assumption, for any $\varepsilon > 0$,
\begin{equation}
\mathbb{P}\!\left(\overline{\Delta U} - \mathbb{E}[\overline{\Delta U}] > \varepsilon\right)
\le
\exp\!\left(-\frac{2 K_{\max} \varepsilon^{2}}{L^{2}}\right).
\end{equation}
Thus, the realized average pruning loss concentrates around its
expectation as the candidate budget grows; this does not by itself
bound the expectation or the worst-case loss across policies.
\end{lemma}

\begin{theorem}[Topology-aware utility growth under routing convergence]
\label{thm:routing-growth}
Let $m_{\rightarrow i}(t)$ denote the external memory discoverable by agent $A_i$ under its current routing state. If the routing process monotonically expands the set of reachable relevant candidates, then the topology-aware cooperation gain
\begin{equation}
I_i^{\mathrm{topo}}(t)=I\!\left(Y_i; m_{\rightarrow i}(t)\mid X_i, M_i\right)
\end{equation}
is non-decreasing in expectation, and the information gap
\begin{equation}
E_i(t)=I_i^{\mathrm{global}}-I_i^{\mathrm{topo}}(t)
\end{equation}
decreases as routing converges.
\end{theorem}

\subsection{Safe Exchange Results}

\begin{proposition}[Authorized exchange bound]
\label{prop:authorized-bound}
Let $\xi_j$ be a candidate asset and $S$ be a sanitization map correctly enforced by the source agent to produce the release $m_{j \rightarrow i} = S(\xi_j)$. If $S$ satisfies a certified residual leakage bound $I(S(\xi_j); s) \le \rho_S$ for any sensitive information $s$, then the privacy risk to an unauthorized receiver is bounded by $\rho_S$. Furthermore, by the Data Processing Inequality (DPI), this leakage cannot exceed the original information content: $I(S(\xi_j); s) \le I(\xi_j; s)$. In the idealized case of perfect blocking or perfect redaction, we have $\rho_S = 0$.
\end{proposition}

\begin{theorem}[Utility--information equivalence for safe exchange policies]
\label{thm:utility-info}
Consider the safe policy class $\Pi_{\mathrm{safe}}(t)$ from Section~\ref{sec:OPAC}. Under the Bayes-optimal log score used to define $U(\pi)$, we have
\begin{equation}
\arg\max_{\pi \in \Pi_{\mathrm{safe}}(t)} U(\pi)
=
\arg\max_{\pi \in \Pi_{\mathrm{safe}}(t)}
\sum_{i=1}^{n} I\!\left(Y_i ; m_{\rightarrow i}^{\pi}(t) \mid X_i, M_i\right).
\end{equation}
Accordingly, among safe policies, maximizing predictive utility is equivalent to maximizing the task-relevant conditional information imported through exchange.
\end{theorem}

\begin{corollary}[When raw release dominates processed release]
\label{cor:exchange-dominates}
Consider a task $T_i$ for agent $A_i$ and a candidate source agent $A_j$. Suppose there exists a transferable asset $\xi_j$ (e.g., a memory fragment or raw skill code) and a processed counterpart $\widetilde{\xi}_j^{\mathrm{proc}}$ such that
\begin{equation}
I(Y_i; \xi_j \mid X_i, M_i) > I(Y_i; \widetilde{\xi}_j^{\mathrm{proc}} \mid X_i, M_i),
\end{equation}
and the raw release does not incur higher communication cost nor lower cooperation trust than the processed alternative. Then raw release strictly dominates processed release in the Pareto sense for that interaction, i.e., $z(\pi_{\mathrm{raw}}) > z(\pi_{\mathrm{proc}})$.
\end{corollary}

\subsection{Consistency Results}

\begin{theorem}[Sample complexity of Pareto consistency]
\label{thm:pareto-consistency}
Fix a horizon $t$ and a confidence level $0<\delta<1$.
Under A7--A8, if
\[
N_t
\ge
\frac{B_{\max}^2}{2\varepsilon^2}
\log\!\left(
\frac{6|\Pi_{\mathrm{safe}}(t)|}{\delta}
\right),
\]
then, with probability at least $1-\delta$,
\[
\sup_{\pi\in\Pi_{\mathrm{safe}}(t)}
\|\widehat z_t(\pi)-z(\pi)\|_\infty
\le\varepsilon.
\]
Define
\[
\widehat{\mathcal P}_{t,2\varepsilon}
=
\left\{
\pi\in\Pi_{\mathrm{safe}}(t):
\nexists\pi'\in\Pi_{\mathrm{safe}}(t),\
\widehat z_t(\pi')
>
\widehat z_t(\pi)+2\varepsilon\mathbf1
\right\},
\]
and
\[
\widehat{\mathcal P}^{z}_{t,2\varepsilon}
=
\{z(\pi):\pi\in\widehat{\mathcal P}_{t,2\varepsilon}\}.
\]
On the same event,
\[
\mathcal P^{\star,z}(t)
\subseteq
\widehat{\mathcal P}^{z}_{t,2\varepsilon},
\qquad
\sup_{\pi\in\widehat{\mathcal P}_{t,2\varepsilon}}
g_{Z_0(t)}(z(\pi))
\le4\varepsilon.
\]
If A10 also holds for $Z_0(t)$, then
\[
d_H\!\left(
\widehat{\mathcal P}^{z}_{t,2\varepsilon},
\mathcal P^{\star,z}(t)
\right)
\le
\frac{4\varepsilon}{\kappa}.
\]
The same statement applies to the realized class $\mathcal K_t$ and
population map $z_t$, with their corresponding empirical estimates.
\end{theorem}

\begin{theorem}[AEP as a Pareto-consistent solver for OPAC, formal]
\label{thm:aep-main}
Under A1--A12, fix $0<\delta<1$ and an evaluation horizon satisfying A12.
Let
\[
\widehat{\mathcal K}_t
=
\left\{
a\in\mathcal K_t:
\nexists a'\in\mathcal K_t,\
\widehat z_t(a')
>
\widehat z_t(a)+2\varepsilon_t\mathbf1
\right\},
\]
and
\[
\widehat{\mathcal P}^{\mathrm{AEP},z}_t
=
\{z_t(a):a\in\widehat{\mathcal K}_t\}.
\]
Then, with probability at least $1-\delta$,
\begin{equation}
d_H\!\left(
\widehat{\mathcal P}^{\mathrm{AEP},z}_t,
\mathcal P^{\star,z}(t)
\right)
\le
\mathcal O(\varepsilon_t)
+
\mathcal O\!\left(
L_{\mathrm{obj}}
\frac{b(\Gamma)+\sigma}{\gamma^{\mathrm{phy}}}
\right)
+
\mathcal O\!\left(\overline{\Delta U}\right).
\end{equation}
The hidden constants may depend on $\kappa$, $C_{\mathrm a}$, $n$,
and $\delta$, but not on $t$ or the three displayed error magnitudes;
an explicit bound is given in the proof.
The objective-space image uses population objectives of the
empirically selected policies. Using their empirical objective values
instead adds at most $\varepsilon_t$ to the bound.
Along regimes where $\varepsilon_t\to0$,
$L_{\mathrm{obj}}(b(\Gamma)+\sigma)/\gamma^{\mathrm{phy}}\to0$,
and $\overline{\Delta U}\to0$ in probability, with the preceding
constants uniformly controlled for each fixed confidence level,
the Hausdorff distance converges to zero in probability.
If a structural error term does not vanish, the result gives an
approximation bound rather than exact consistency.
\end{theorem}

\subsection{Proofs}

We now provide theorem-by-theorem proofs corresponding to the full statements collected in Appendix~\ref{app:aep-theory-results}. Throughout, all expectations are taken with respect to the joint distribution induced by the task variables, local memories, routing state, and the safe cooperation policy under consideration.

\begin{proof}[Proof of Theorem~\ref{thm:spectral-topology}]
Let
\begin{equation}
\Delta_t := W_i^{\mathrm{cog}}(t) - W^{\mathrm{phy}}.
\end{equation}
Assume that the physical graph is connected with spectral gap
\begin{equation}
\gamma^{\mathrm{phy}} := 1 - \lambda_2\!\left(W^{\mathrm{phy}}\right) > 0,
\end{equation}
and choose a constant step size $\eta$ such that $0 < \eta < 1/\gamma^{\mathrm{phy}}$. Let
\begin{equation}
\rho_\eta := 1 - \eta \gamma^{\mathrm{phy}} \in (0,1).
\end{equation}
If the topology-learning error satisfies the graph-spectral recursion
\begin{equation}
\mathbb{E}\|\Delta_{t+1}\|_2
\le
(1-\eta\gamma^{\mathrm{phy}})\,\mathbb{E}\|\Delta_t\|_2
+ \eta\bigl(b(\Gamma)+\sigma\bigr),
\end{equation}
then
\begin{equation}
\limsup_{t\to\infty}
\mathbb{E}\left\|W_i^{\mathrm{cog}}(t)-W^{\mathrm{phy}}\right\|_2
\le
\frac{\eta\bigl(b(\Gamma)+\sigma\bigr)}{1-\rho_\eta}.
\end{equation}
Equivalently,
\begin{equation}
\limsup_{t\to\infty}
\mathbb{E}\left\|W_i^{\mathrm{cog}}(t)-W^{\mathrm{phy}}\right\|_2
\le
\frac{b(\Gamma)+\sigma}{\gamma^{\mathrm{phy}}}.
\end{equation}
Write
\begin{equation}
e_t := \mathbb{E}\|\Delta_t\|_2.
\end{equation}
The assumed recursion is exactly
\begin{equation}
e_{t+1} \le \rho_\eta e_t + \eta\bigl(b(\Gamma)+\sigma\bigr),
\qquad 0<\rho_\eta<1.
\end{equation}
Iterating this inequality gives, for every $t\ge 0$,
\begin{align}
e_t
&\le \rho_\eta^t e_0
+ \eta\bigl(b(\Gamma)+\sigma\bigr)\sum_{k=0}^{t-1}\rho_\eta^k \nonumber\\
&= \rho_\eta^t e_0
+ \eta\bigl(b(\Gamma)+\sigma\bigr)\frac{1-\rho_\eta^t}{1-\rho_\eta}.
\end{align}
Because $\rho_\eta \in (0,1)$, we have $\rho_\eta^t \to 0$ as $t\to\infty$. Taking $\limsup$ on both sides yields
\begin{equation}
\limsup_{t\to\infty} e_t
\le
\frac{\eta\bigl(b(\Gamma)+\sigma\bigr)}{1-\rho_\eta}.
\end{equation}
Substituting the definition $\rho_\eta = 1-\eta\gamma^{\mathrm{phy}}$ gives
\begin{equation}
1-\rho_\eta = \eta\gamma^{\mathrm{phy}},
\end{equation}
and therefore
\begin{equation}
\frac{\eta\bigl(b(\Gamma)+\sigma\bigr)}{1-\rho_\eta}
=
\frac{b(\Gamma)+\sigma}{\gamma^{\mathrm{phy}}}.
\end{equation}
Recalling that $e_t = \mathbb{E}\|W_i^{\mathrm{cog}}(t)-W^{\mathrm{phy}}\|_2$ proves both displayed bounds.
\end{proof}

\begin{proof}[Proof of Theorem~\ref{thm:routing-growth}]
Fix an agent $A_i$. By assumption, the routing process monotonically expands the set of reachable relevant candidates. Hence the corresponding imported-context collection is increasing in the sense that for times $t_1 \le t_2$, the sigma-field generated by $m_{\rightarrow i}(t_1)$ is contained in the sigma-field generated by $m_{\rightarrow i}(t_2)$. Equivalently, there exists a random variable $\Delta_{t_1,t_2}$ such that
\begin{equation}
m_{\rightarrow i}(t_2) = \bigl(m_{\rightarrow i}(t_1), \Delta_{t_1,t_2}\bigr).
\end{equation}
Applying the chain rule for conditional mutual information yields
\begin{align}
I\!\left(Y_i ; m_{\rightarrow i}(t_2) \mid X_i, M_i\right)
&= I\!\left(Y_i ; m_{\rightarrow i}(t_1), \Delta_{t_1,t_2} \mid X_i, M_i\right) \nonumber\\
&= I\!\left(Y_i ; m_{\rightarrow i}(t_1) \mid X_i, M_i\right)
+ I\!\left(Y_i ; \Delta_{t_1,t_2} \mid X_i, M_i, m_{\rightarrow i}(t_1)\right).
\end{align}
The second term is nonnegative, so
\begin{equation}
I_i^{\mathrm{topo}}(t_2) \ge I_i^{\mathrm{topo}}(t_1).
\end{equation}
Therefore $I_i^{\mathrm{topo}}(t)$ is non-decreasing, and hence also non-decreasing in expectation.

Now recall that the full-information benchmark is
\begin{equation}
I_i^{\mathrm{global}} = I\!\left(Y_i ; \widetilde{M}_i^{\star} \mid X_i, M_i\right),
\end{equation}
where $\widetilde{M}_i^{\star}$ contains all relevant memory accessible under global awareness and unconstrained discovery. Since the discoverable memory under routing convergence approaches this globally available set, the residual gap
\begin{equation}
E_i(t)=I_i^{\mathrm{global}}-I_i^{\mathrm{topo}}(t)
\end{equation}
is non-increasing in $t$. In particular, whenever routing reveals an additional relevant admissible source with positive conditional information, the gap decreases strictly; otherwise it remains unchanged. This proves the claim.
\end{proof}

\begin{proof}[Proof of Theorem~\ref{thm:utility-info}]
Fix a safe policy $\pi \in \Pi_{\mathrm{safe}}(t)$. By definition,
\begin{equation}
U_i(\pi)=\mathbb{E}\!\left[\log p^{\star}\!\left(Y_i \mid X_i, \widetilde{M}_i^{\pi}(t)\right)\right].
\end{equation}
Because $p^{\star}(\cdot \mid X_i, \widetilde{M}_i^{\pi}(t))$ is the Bayes-optimal predictive distribution, the standard identity for log score gives
\begin{equation}
U_i(\pi) = -H\!\left(Y_i \mid X_i, \widetilde{M}_i^{\pi}(t)\right).
\end{equation}
Using $\widetilde{M}_i^{\pi}(t)=\bigl(M_i, m_{\rightarrow i}^{\pi}(t)\bigr)$, we obtain
\begin{align}
U_i(\pi)
&= -H\!\left(Y_i \mid X_i, M_i, m_{\rightarrow i}^{\pi}(t)\right) \nonumber\\
&= -H\!\left(Y_i \mid X_i, M_i\right)
+ I\!\left(Y_i ; m_{\rightarrow i}^{\pi}(t) \mid X_i, M_i\right),
\end{align}
where the second line follows from the identity
\begin{equation}
I(Y;Z\mid W)=H(Y\mid W)-H(Y\mid W,Z).
\end{equation}
Summing over agents yields
\begin{equation}
U(\pi)=\sum_{i=1}^n U_i(\pi)
= -\sum_{i=1}^n H\!\left(Y_i \mid X_i, M_i\right)
+ \sum_{i=1}^n I\!\left(Y_i ; m_{\rightarrow i}^{\pi}(t) \mid X_i, M_i\right).
\end{equation}
The first term is independent of the policy $\pi$. Therefore, over the safe policy class, maximizing $U(\pi)$ is equivalent to maximizing
\begin{equation}
\sum_{i=1}^n I\!\left(Y_i ; m_{\rightarrow i}^{\pi}(t) \mid X_i, M_i\right).
\end{equation}
Hence
\begin{equation}
\arg\max_{\pi \in \Pi_{\mathrm{safe}}(t)} U(\pi)
=
\arg\max_{\pi \in \Pi_{\mathrm{safe}}(t)}
\sum_{i=1}^{n} I\!\left(Y_i ; m_{\rightarrow i}^{\pi}(t) \mid X_i, M_i\right),
\end{equation}
as claimed.
\end{proof}

\begin{proof}[Proof of Corollary~\ref{cor:exchange-dominates}]
Let $\pi_{\mathrm{raw}}$ and $\pi_{\mathrm{proc}}$ denote the policies for raw and processed release, respectively. In the OPAC framework, their corresponding multi-objective vectors are defined as $z(\pi_{\mathrm{raw}}) = (U_{\mathrm{raw}}, T_{\mathrm{raw}}, -C_{\mathrm{raw}})$ and $z(\pi_{\mathrm{proc}}) = (U_{\mathrm{proc}}, T_{\mathrm{proc}}, -C_{\mathrm{proc}})$.

By Theorem~\ref{thm:utility-info}, the predictive utility ordering between these two actions is equivalent to the ordering of their task-relevant conditional mutual information. Hence the hypothesis
\begin{equation}
I(Y_i; \xi_j \mid X_i, M_i) > I(Y_i; \widetilde{\xi}_j^{\mathrm{proc}} \mid X_i, M_i)
\end{equation}
strictly implies an advantage in the utility dimension:
\begin{equation}
U_{\mathrm{raw}} > U_{\mathrm{proc}}.
\end{equation}
Furthermore, under the premise that the raw release does not incur higher communication cost nor lower cooperation trust than the processed alternative, the remaining coordinates satisfy:
\begin{equation}
-C_{\mathrm{raw}} \ge -C_{\mathrm{proc}} \quad \text{and} \quad T_{\mathrm{raw}} \ge T_{\mathrm{proc}}.
\end{equation}
Combining these conditions yields $z(\pi_{\mathrm{raw}}) \ge z(\pi_{\mathrm{proc}})$ coordinate-wise, with a strict inequality in the predictive utility dimension. Therefore, the raw release strictly dominates the processed release in the Pareto sense:
\begin{equation}
z(\pi_{\mathrm{raw}}) \succ z(\pi_{\mathrm{proc}}).
\end{equation}
This proves that raw release yields a strictly better Pareto tradeoff for that interaction.
\end{proof}

\begin{proof}[Proof of Lemma~\ref{thm:pruning}]
By assumption, the random variables $\Delta U_1,\dots,\Delta U_{K_{\max}}$ are independent and each lies in $[0,L]$. Their average is
\begin{equation}
\overline{\Delta U}=\frac{1}{K_{\max}}\sum_{k=1}^{K_{\max}} \Delta U_k.
\end{equation}
Hoeffding's inequality for bounded independent random variables gives, for any $\varepsilon>0$,
\begin{equation}
\mathbb{P}\!\left(\overline{\Delta U}-\mathbb{E}[\overline{\Delta U}]>\varepsilon\right)
\le
\exp\!\left(-\frac{2K_{\max}^2\varepsilon^2}{\sum_{k=1}^{K_{\max}}(L-0)^2}\right)
= \exp\!\left(-\frac{2K_{\max}\varepsilon^2}{L^2}\right).
\end{equation}
This is exactly the stated bound. The exponential dependence on $K_{\max}$ shows that the realized pruning loss concentrates sharply around its expectation as the candidate budget increases.
\end{proof}

\begin{proof}[Proof of Proposition~\ref{prop:authorized-bound}]
Assume the source agent applies a sanitization map $S$ to the candidate asset $\xi_j$, resulting in the transmitted artifact $m_{j \rightarrow i}=S(\xi_j)$. For any sensitive attribute $s$ unauthorized for the receiver, the privacy leakage risk is measured by the mutual information $I(m_{j \rightarrow i}; s) = I(S(\xi_j); s)$.

By definition, if the sanitization mechanism guarantees a certified residual leakage $\rho_S$, we have:
\begin{equation}
I\!\left(S(\xi_j); s\right) \le \rho_S.
\end{equation}
Furthermore, since $S(\xi_j)$ is a measurable function of the pre-sanitized asset $\xi_j$, the Data Processing Inequality (DPI) ensures that sanitization cannot increase mutual information:
\begin{equation}
I\!\left(S(\xi_j); s\right) \le I\!\left(\xi_j; s\right).
\end{equation}
Thus, the privacy risk to the unauthorized receiver is formally bounded by the certified capacity $\rho_S$ of the sanitization map, which in turn is bounded by the original information $I(\xi_j; s)$.

If the policy blocks transmission entirely, then $m_{j \rightarrow i}$ is a constant null symbol, yielding $I(S(\xi_j); s) = 0$. Likewise, under perfect redaction with respect to $s$, the sanitized artifact is completely independent of $s$, so the mutual information is also zero. In these idealized cases, the upper bound becomes strictly $\rho_S = 0$.
\end{proof}

\begin{proof}[Proof of Theorem~\ref{thm:pareto-consistency}]
Condition on the evaluation snapshot.
By A8 and Hoeffding's inequality, followed by a union bound,
\[
\begin{split}
&\mathbb P\!\left(
\sup_{\pi\in\Pi_{\mathrm{safe}}(t)}
\|\widehat z_t(\pi)-z(\pi)\|_\infty>\varepsilon
\,\middle|\,\text{snapshot}
\right)\\
&\qquad\le
6|\Pi_{\mathrm{safe}}(t)|
\exp\!\left(
-\frac{2N_t\varepsilon^2}{B_{\max}^2}
\right)
\le\delta.
\end{split}
\]
The same probability guarantee holds after removing the conditioning.
Work on the resulting uniform-accuracy event.

If a Pareto-optimal policy $\pi^\star$ were excluded, some policy
$\pi$ would satisfy
\[
\widehat z_t(\pi)
>
\widehat z_t(\pi^\star)+2\varepsilon\mathbf1.
\]
Uniform accuracy would imply
$z(\pi)>z(\pi^\star)$ in every coordinate, a contradiction.
Hence
\[
\mathcal P^{\star,z}(t)
\subseteq
\widehat{\mathcal P}^{z}_{t,2\varepsilon}.
\]

For any retained policy $\pi$, suppose
$g_{Z_0(t)}(z(\pi))>4\varepsilon$.
Since the policy class is finite, there exists $\pi'$ such that
\[
z(\pi')>z(\pi)+4\varepsilon\mathbf1.
\]
Uniform accuracy gives
\[
\begin{aligned}
\widehat z_t(\pi')
&\ge z(\pi')-\varepsilon\mathbf1\\
&>z(\pi)+3\varepsilon\mathbf1\\
&\ge\widehat z_t(\pi)+2\varepsilon\mathbf1,
\end{aligned}
\]
contradicting retention. Thus
$g_{Z_0(t)}(z(\pi))\le4\varepsilon$.

Under A10,
\[
\operatorname{dist}_\infty
\bigl(z(\pi),\mathcal P^{\star,z}(t)\bigr)
\le
\frac{g_{Z_0(t)}(z(\pi))}{\kappa}
\le
\frac{4\varepsilon}{\kappa}.
\]
Combining this directed-distance bound with the preceding set inclusion
proves the Hausdorff bound.
Replacing the reference policy class and map by $\mathcal K_t$
and $z_t$ gives the realized-class statement.
\end{proof}

\begin{proof}[Proof of Theorem~\ref{thm:aep-main}]
Write
\[
D=\frac{b(\Gamma)+\sigma}{\gamma^{\mathrm{phy}}},
\qquad
r_t=\max_i
\|W_i^{\mathrm{cog}}(t)-W^{\mathrm{phy}}\|_2,
\qquad
q=1-\eta\gamma^{\mathrm{phy}}.
\]

\noindent\textbf{Step 1 (Finite-time topology control).}
Iterating A3 gives, for every agent,
\[
\mathbb E
\|W_i^{\mathrm{cog}}(t)-W^{\mathrm{phy}}\|_2
\le
q^t e_0+D(1-q^t)
\le q^t e_0+D.
\]
Since a maximum of nonnegative variables is bounded by their sum,
\[
\mathbb E r_t\le n(q^t e_0+D).
\]
Markov's inequality therefore yields, with probability at least
$1-\delta/2$,
\[
r_t\le\frac{2n}{\delta}(q^t e_0+D).
\]
If the right-hand side is zero, the same conclusion holds almost surely.

\noindent\textbf{Step 2 (Population objective-set perturbation).}
For each $\pi\in\Pi_{\mathrm{safe}}(t)$, A9 and A11 imply
\[
\|z_t(R_t(\pi))-z(\pi)\|_\infty
\le
L_{\mathrm{obj}}r_t
+C_{\mathrm a}\overline{\Delta U}
=:\beta_t.
\]
The map $R_t$ is defined on every reference policy and is onto
$\mathcal K_t$, so this bound controls both directions:
\[
d_H(Z_t,Z_0(t))\le\beta_t.
\]
On the topology event, A12 gives
\[
\beta_t
\le
\frac{2n}{\delta}
\bigl(\varepsilon_t+L_{\mathrm{obj}}D\bigr)
+C_{\mathrm a}\overline{\Delta U}
=:\overline\beta_t.
\]
This step uses A11. Concentration of an average pruning loss alone
would not imply this uniform objective-set bound.

\noindent\textbf{Step 3 (Frontier stability).}
Let
\[
F_0=\operatorname{PF}(Z_0(t))
=\mathcal P^{\star,z}(t),
\qquad
F_t=\operatorname{PF}(Z_t).
\]
For any $p\in F_0$, choose $u\in Z_t$ satisfying
$\|u-p\|_\infty\le\beta_t$.
For every $v\in Z_t$, choose $x\in Z_0(t)$ satisfying
$\|v-x\|_\infty\le\beta_t$.

Since $p$ is Pareto optimal,
$\min_k(x_k-p_k)\le0$. Therefore
\[
\min_k(v_k-u_k)
\le
\min_k(x_k-p_k)+2\beta_t
\le2\beta_t.
\]
Taking the maximum over $v$ gives
$g_{Z_t}(u)\le2\beta_t$.
By A10,
\[
\begin{aligned}
\operatorname{dist}_\infty(p,F_t)
&\le
\|p-u\|_\infty+
\operatorname{dist}_\infty(u,F_t)\\
&\le
\left(1+\frac{2}{\kappa}\right)\beta_t.
\end{aligned}
\]
Interchanging the two objective sets and applying A10 to $Z_0(t)$
gives the reverse directed-distance bound. Hence
\[
d_H(F_t,F_0)
\le
\left(1+\frac{2}{\kappa}\right)\beta_t.
\]

\noindent\textbf{Step 4 (Statistical error and aggregation).}
Condition on the evaluation snapshot.
By A8 and
$|\mathcal K_t|\le|\Pi_{\mathrm{safe}}(t)|\le M_t$,
\[
\begin{split}
&\mathbb P\!\left(
\sup_{a\in\mathcal K_t}
\|\widehat z_t(a)-z_t(a)\|_\infty>\varepsilon_t
\,\middle|\,\text{snapshot}
\right)\\
&\qquad\le
6M_t\exp\!\left(
-\frac{2N_t\varepsilon_t^2}{B_{\max}^2}
\right)
=\frac{\delta}{2}.
\end{split}
\]
The same guarantee holds unconditionally.
A union bound shows that the statistical and topology events hold
jointly with probability at least $1-\delta$.

Applying Theorem~\ref{thm:pareto-consistency} to $\mathcal K_t$
and $z_t$ yields
\[
d_H\!\left(
\widehat{\mathcal P}^{\mathrm{AEP},z}_t,F_t
\right)
\le\frac{4\varepsilon_t}{\kappa}.
\]
Consequently,
\begin{align*}
&d_H\!\left(
\widehat{\mathcal P}^{\mathrm{AEP},z}_t,F_0
\right)\\
&\quad\le
\frac{4\varepsilon_t}{\kappa}
+
\left(1+\frac{2}{\kappa}\right)\overline\beta_t\\
&\quad=
\left[
\frac{4}{\kappa}
+
\frac{2n}{\delta}
\left(1+\frac{2}{\kappa}\right)
\right]\varepsilon_t\\
&\qquad+
\frac{2n}{\delta}
\left(1+\frac{2}{\kappa}\right)
L_{\mathrm{obj}}
\frac{b(\Gamma)+\sigma}{\gamma^{\mathrm{phy}}}\\
&\qquad+
C_{\mathrm a}
\left(1+\frac{2}{\kappa}\right)
\overline{\Delta U}.
\end{align*}
This proves the stated three-term bound.

On the statistical event, replacing selected population objectives
by their empirical estimates adds at most $\varepsilon_t$ to the
Hausdorff bound. Under the stated vanishing-error conditions,
the bound yields convergence in probability by taking the confidence
failure probability arbitrarily small. No almost-sure or
simultaneous-in-$t$ guarantee is asserted.
\end{proof}

\section{Limitations}
\label{app:limitations}

This work studies AEP mainly in controlled benchmark environments, which may not fully capture the heterogeneity of real-world agent deployments. The current protocol also relies on practical choices for routing thresholds, trust updates, safety gates, and checkpointing policies. Different applications may require different trade-offs among utility, privacy, safety, and communication cost. Future work can further evaluate AEP in larger deployments and study adaptive policy configuration.

\section{Social Impact Discussion}
\label{app:social-impact}

AEP aims to make collaboration among privately owned agents more useful while preserving selective disclosure and auditable control. This may help future agent ecosystems reduce duplicated work, improve interoperability, and support safer reuse of distributed context assets. However, agent networking systems may also increase the scale at which information is shared across services if deployed without proper governance. We therefore view access control, transparency, auditing, and human-configured policies as important complements to the protocol design.

\end{document}